\documentclass{aa}  

\usepackage[super]{nth}
\usepackage{graphicx}
\usepackage{multirow}
\usepackage[separate-uncertainty=true, retain-unity-mantissa = false]{siunitx}

\DeclareSIUnit\angstrom{\text {Å}}
\DeclareUnicodeCharacter{0301}{\'{e}}

\newcommand{\figref}[1]{\figurename~\ref{#1}}
\newcommand{\secref}[1]{Sect.~\ref{#1}}
\newcommand{\tabref}[1]{\tablename~\ref{#1}}

\newcommand{\ptre}{$\mathrm{PAH_{3.3}}\ $}

\usepackage[varg]{txfonts}

\usepackage{xcolor}
\definecolor{blue_dark}{rgb}{0.2, 0.2, 0.6}
\definecolor{blue_green}{rgb}{0.1, 0.4, 0.3}
\usepackage[hidelinks,colorlinks=true,linkcolor={blue_dark},citecolor={blue_dark}, urlcolor={blue_green},filecolor={blue_dark}]{hyperref}
\usepackage{orcidlink}

\begin{document} 

   \title{Spatially resolved $\mathrm{PAH_{3.3}}$ emission and stellar ages in ram pressure stripped clumps at $z\sim0.3$}

   \author{Pietro Benotto\orcidlink{0009-0004-7602-0160}\inst{1,2}\fnmsep\thanks{\email{pietro.benotto@inaf.it}}
   \and
   Benedetta Vulcani\orcidlink{0000-0003-0980-1499}\inst{1}
   \and
   Peter J. Watson\orcidlink{0000-0003-3108-0624}\inst{1}
   \and
   Giulia Rodighiero\orcidlink{0000-0002-9415-2296}\inst{2}
   \and
   Bianca M. Poggianti\orcidlink{0000-0001-8751-8360}\inst{1}
   \and
   Marco Gullieuszik\orcidlink{0000-0002-7296-9780}\inst{1}
   \and
   Jacopo Fritz\orcidlink{0000-0002-7042-1965}\inst{3}
   \and
   Thomas S.-Y. Lai\orcidlink{0000-0001-8490-6632}\inst{4}
   \and
   Augusto E. Lassen\orcidlink{0000-0003-3575-8316}\inst{1}
   \and
   Matthew A. Malkan\orcidlink{0000-0001-6919-1237}\inst{5}
   \and
   Alessia Moretti\orcidlink{0000-0002-1688-482X}\inst{1}
   }

   \institute{
        \inst{1} INAF, Osservatorio Astronomico di Padova, Vicolo dell’Osservatorio 5, 35122, Padova, Italy.\\
        \inst{2} Dipartimento di Fisica e Astronomia `G. Galilei', Università di Padova, Vicolo dell'Osservatorio 3, 35122 Padova, Italy.\\
        \inst{3} Instituto de Radioastronomia y Astrofisica, UNAM, Campus Morelia, A.P. 3-72, C.P. 58089, Morelia, Mexico.\\
        \inst{4} IPAC, California Institute of Technology, 1200 E. California Blvd., Pasadena, CA 91125, USA.\\
        \inst{5} University of California, Los Angeles, Department of Physics and Astronomy, 430 Portola Plaza, Los Angeles, CA 90095, USA.
    }
   
   \date{}
    
   \abstract{Ram pressure stripping (RPS) plays a crucial role in shaping galaxy evolution in dense environments, yet its impact on the molecular and dusty phases of the interstellar medium remains poorly understood. We present JWST/NIRCam $\SI{3.3}{\micro\meter}$ polycyclic aromatic hydrocarbon (PAH) emission maps for the nine most striking RPS galaxies in the Abell 2744 cluster at redshift $z_{cl}=0.306$, tracing the effects of environmental processes on small dust grains. Exploiting multi-band JWST/NIRCam and HST photometry, we performed a spatially resolved ultraviolet (UV) to mid-infrared (MIR)\ spectral energy distribution (SED) fitting to characterise stellar populations in both galactic disks and clumps detected in the stripped tails.
   We detected \ptre emission in eight of the nine galaxies at $5\sigma$, with morphologies revealing disk truncation and elongation along the RPS direction. In three galaxies, \ptre emission is also found in star-forming clumps embedded in the stripped tails up to a distance of 40~kpc. Star formation rates inferred from \ptre emission are in agreement with those derived from SED fitting averaged over the past $\SI{100}{Myr}$ within an intrinsic scatter of $\SI{0.4}{dex}$, but the relation appears to be age-dependent. The spatial correlation between the PAH strength, stellar age, and  star formation rate (SFR) is consistent across disks and tails and demonstrates that PAH-carrying molecules can survive and become stripped by ram pressure. Finally, age gradients revealed by the SED fitting provide observational evidence of the fireball model in star-forming, stripped clumps of galaxies at $z \sim 0.3$. This work represents the first detailed study of PAH emission in cluster galaxies, offering new insights into the fate of dust and star formation in extreme environments.}
   
   \authorrunning{P. Benotto et al.}
   \maketitle

\section{Introduction}
The mid-infrared (MIR) coverage of the James Webb Space Telescope (JWST; \citealt{Gardner:2006})
 and its unprecedented sensitivity and resolution are transforming our understanding of interstellar polycyclic aromatic hydrocarbons (PAHs).
The ubiquitous presence of these very large molecules of aromatic rings in the interstellar medium (ISM) is traced by emission features ranging from $\SI{3}{\micro\meter}$ to $\SI{20}{\micro\meter}$. Thanks to their ability to absorb UV photons and re-emit them in MIR bands, they can be used to trace star formation in dusty environments \citep{Forster:2004, Peeters:2004, Shipley:2016, Xie:2019}. 
In addition, PAHs play an active role in the thermal balance of the ISM through photoelectric heating, influencing the structure of photodissociation regions and the formation of molecular clouds \citep[e.g.][]{Bakes:1994, Hollenbach:1999, Weingartner:2001, Tielens:2008}. Their abundance and spectral properties reflect the life cycle of interstellar dust, from stellar injection from asymptotic giant branch (AGB) stellar winds and supernova ejecta to processing in shocks. Indeed, the intensity of the emission in each band depends on the properties of the emitting molecules, such as size and charge state \citep{Allamandola:1985, Allamandola:1999, Tielens:2008}, as well as on the physical properties of the galaxy environment; whereas the presence of shocks or hard photons can destroy these molecules \citep{Micelotta:2010, Slavin:2015, Geballe:1989, Tielens:2008, Herrero:2014, Rigopoulou:2024}. 

Before the launch of JWST, our understanding of PAHs was primarily based on the AKARI mission \citep{Murakami:2007} and the Spitzer Space Telescope \citep{Houk:2004}. Initial research was focussed on identifying MIR features as vibrational modes of PAH molecules and calibrating the $\SI{5}{\micro\meter}$ to $\SI{20}{\micro\meter}$ bands as indicators of star formation \citep{Leger:1984, Allamandola:1985, Tielens:2008, Peeters:2004, Forster:2004}. It was also revealed that the strength and profile of the PAH bands vary with the hardness of the local radiation field, the ionisation state of the carriers, and the metallicity of the host galaxy, with a deficit of PAH emission in low-metallicity ($Z\lesssim\SI{0.25}{Z_\odot}$) systems \citep[e.g.][]{Engelbracht:2005, Calzetti:2007, Draine:2007, Smith:2007}. Additionally, it was noted that these molecules are primarily excited by radiation from B-type stars and the photodissociation regions around O-type stars; however, they do not directly survive within \textsc{H\;\!ii} regions \citep{Helou:2004, Bendo:2006}. For these reasons, PAHs effectively trace star formation over the last 100 million years, similar to far-IR (FIR) radiation \citep{Calzetti:2007, Kennicutt:2012}. However, the relatively low sensitivity and angular resolution of pre-JWST facilities have restricted detailed investigations to the brightest nearby galaxies and star-forming regions, with only a few studies extending beyond the Local Universe, which have mostly targeted luminous IR galaxies and lack spatially resolved information \citep[e.g.][]{Yan:2005, Takagi:2010, Murata:2014}. 

Among the PAH bands, the $\SI{3.3}{\micro\meter}$ one was the least studied in the pre-JWST era, both because of its intrinsic weakness and because it falls near a water atmospheric absorption band, limiting its observation in the Local Universe from the ground. Observations of galactic regions first established the 3.3 µm band as the C–H stretching mode of small, mostly neutral PAHs \citep{Tokunaga:1991, vanDiedenhoven:2004, Tielens:2008}. Space IR telescopes, starting with the Infrared Space Observatory (ISO) and on to AKARI, have extended these studies to nearby star-forming galaxies and active nuclei, showing that the $\SI{3.3}{\micro\meter}$ emission is typically an order of magnitude weaker than the PAH bands at longer wavelengths and is suppressed in harsh radiation fields and low-metallicity environments \citep[e.g.][]{Moorwood:1986, Imanishi:2008, Murata:2014, Lai:2020}. 

With the advent of JWST, the \ptre emission has become an essential observational diagnostic. As the only significant PAH band covered by both NIRCam and NIRSpec, its study is greatly enhanced by the high spatial resolution of these instruments. PHANGS–JWST imaging provided maps of nearby galaxies, with a resolution smaller than $\SI{10}{pc}$, of the $\SI{3.3}{\micro\meter}$ emission, findind its intensity to be tightly correlated with the $\SI{17}{\micro\meter}$ one, confirming the link of the \ptre band to small, mostly neutral PAHs \citep{Sandstrom2023, Chastenet2023}. The spectroscopy of luminous IR galaxies showed enhanced \ptre emission in circumnuclear star-forming rings and suppression near AGN, indicating the vulnerability of small PAHs to hard radiation fields \citep{Lai:2023}. The \ptre band has also been calibrated as a star formation rate (SFR) indicator \citep{Lai:2020, Gregg:2024}. At higher redshift, the \ptre band has been extensively observed in dusty star-forming galaxies, correlating with the $\mathrm{H\alpha}$ and IR emission \citep{Vulcani:2023, Vulcani:2025, Lyu:2024, Cheng:2025, Shivaei:2024}, strengthening its ability to trace the recent star formation.

Studying the \ptre feature in galaxy clusters offers a unique opportunity to probe the survival and evolution of small, neutral PAH molecules in the harsh conditions that characterise these extreme environments. With JWST, we are now able to obtain spatially resolved maps of the \ptre band in cluster galaxies, enabling detailed investigations of how environmental processes, such as ram-pressure stripping (RPS), intra-cluster radiation fields and shocks, affect the PAH population. In this work, we focus on RPS galaxies, examining the strength and spatial distribution of the \ptre emission as a tracer of dust processing under extreme environmental conditions. Overall, RPS occurs when the intracluster medium (ICM) exerts pressure on the ISM of a galaxy, stripping gas from the galaxy as it moves through the cluster \citep{Gunn:1972}.

Numerous studies have shown that many RPS galaxies host significant star formation in their stripped tails, extending well beyond their stellar disks. This star formation is traced by ionised gas and young stellar populations \citep{Poggianti:2017, Gullieuszik:2017, Bellhouse:2019, Moretti:2020} and it is thought to arise from the in situ cooling and condensation of molecular gas,  forming giant molecular clouds outside the galaxy stellar body  \citep{Moretti:2020, Moretti:2023}. Observations of CO emission confirm the presence of cold molecular gas in these tails \citep{Jachym:2014, Jachym:2017, Moretti:2023}. 
The morphology of these star-forming regions, namely, made up of compact knots embedded in elongated tails \citep{Poggianti:2019}, has led to the development of the fireball model \citep{Yoshida:2008, Kenney:2014}, which explains their structure as a result of ongoing RPS.  In this scenario, star-forming clumps become decoupled from the surrounding gas, which continues to be pushed downstream by ram pressure. Further collapse can occur later, leading to younger stellar populations forming at increasing distances from the galaxy \citep{Yoshida:2008, Kenney:2014,  Gullieuszik:2023, Werle:2024}. So far, observational evidence for the fireball model has been limited to clusters in the Local Universe ($z<0.07$). Overall, RPS contributes to galaxy quenching by removing the gas reservoir, but not before triggering star formation both within disks and in extraplanar regions \citep{Poggianti:2017, Vulcani:2018}. 

In the Local Universe, previous studies have shown that PAHs can survive in the tail of RPS systems. For instance, using Spitzer/IRAC photometry, \citet{Kenney:2008} found PAH emission in extraplanar regions of RPS galaxies in the Virgo cluster, likely excited by local star formation or interactions with the ICM. Similarly, using Spitzer/IRS and IRAC, \citet{Sivanandam:2014}  detected PAHs associated with shocked molecular gas in the RPS tail of a galaxy in Abell 3627, suggesting that shocks influence both PAH excitation and distribution. These findings support the use of PAHs as tracers of the ISM in cluster environments. 

At intermediate redshift, \citet{Vulcani:2025} used NIRSpec spectroscopy to show that galaxies in Abell 2744 cluster with high \ptre equivalent width often display morphological disturbances consistent with RPS activity. However, the spectroscopic data lacked the spatial resolution to map the distribution of the \ptre emission. The potential for detecting \ptre emission maps from NIRCam photometry in the Abell 2744 cluster has been proposed in the Megascience data release \citep{Suess:2024}. Working on this photometry, \citet{Cheng:2025} reported statistical detections of \ptre emission in the cluster, establishing a connection between PAH strength and recent peaks in star formation history (SFH).
While not focussed on RPS galaxies, they noted qualitative asymmetries in the PAH maps of some stripped systems, suggesting environmental effects may influence PAH morphology and emphasising the significance of \ptre emission for identifying dusty star-forming galaxies.

In this work, we undertook a detailed analysis of a sample of RPS galaxies at $z\sim0.3$  by exploiting the unprecedented capabilities of HST and, in particular, JWST. These new observations allowed us to investigate aspects of RPS systems that were previously inaccessible. 

Our study follows two complementary approaches: (i) we analysed the spatially resolved stellar population ages in the stripped tails to test whether the fireball model is applicable at this epoch; and (ii) we characterise the distribution and strength of the PAH emission to assess the survival of PAH molecules in the tail environment as well as to investigate how their emission is correlated with key physical quantities, such as the SFR, compared to the galactic disk. This enables us to evaluate whether PAH emission in extraplanar regions is still primarily powered by in situ star formation. By combining these two lines of investigation, we directly connect PAH properties with the recent star-formation history, providing a level of insight into RPS tails that has only become possible with JWST.

To achieve this, we explored the spatial distribution of \ptre emission in RPS galaxies within the Abell 2744 cluster using NIRCam photometry. We focussed on both the disturbed galactic disks and the star-forming clumps in the stripped tails, simultaneously extracting PAH emission maps and performing spatially resolved spectral energy distribution (SED) fitting across each galaxy. In \secref{sec:sample}, we describe the sample selection, and the available multi-wavelength data. In \secref{sec:methods}, we outline the data analysis and methodology. In \secref{sec:results}, we examine the effectiveness of \ptre emission as a star formation tracer and we characterise the stellar age of the clumps in the stripped tails. We discuss the implications of our findings in \secref{sec:discussion} and summarise our conclusions in \secref{sec:conclusions}. A standard $\Lambda$CDM cosmology with $\Omega_\mathrm{m}=0.3$, $\Omega_\mathrm{\Lambda}=0.7$, and $\mathrm{H_0}=\SI{70}{km\,s^{-1}\,Mpc^{-1}}$ is adopted throughout.

\section{Sample presentation}\label{sec:sample}

\begin{table*}[htbp]
    \caption{\centering Properties of the sample}
    \label{tab:sample}
    \centering
    \begin{tabular}{ccccccccc}
    \hline\hline
         ID & RA [hms] & DEC [dms] & $z_\mathrm{spec}$ & $\log_{10}(M_\star/\mathrm{M_\odot})$ & UV coverage & RPS identification & $z_\mathrm{spec}$ reference\\
         \hline
         8783 & 00:14:19.9 & -30:23:59 & 0.291 & 8.67 & Yes &\citet{Moretti:2022}&\citet{Bergamini:2023}\\
         12875 & 00:14:19.4 & -30:23:27 & 0.294 & 10.12 & Yes & \citet{Moretti:2022}&\citet{Merlin:2024}\\
         13821 & 00:14:22.4 & -30:23:03 & 0.296 & 10.44 & Yes & \citet{Moretti:2022}&\citet{Merlin:2024}\\
         14826 & 00:14:20.1 & -30:23:04 & 0.303 & 9.42 & Yes & \citet{Moretti:2022}&\citet{Merlin:2024}\\
         10685 & 00:14:26.6 & -30:23:44 & 0.303 & 9.72 & Yes & \citet{Watson:2025}  &\citet{Merlin:2024}\\
         11531 & 00:14:28.5 & -30:23:34 & 0.302 & 9.38 & >$\SI{330}{nm}$ & \citet{Watson:2025}&\citet{Merlin:2024}\\
         9856 & 00:14:25.7 & -30:24:12 & 0.297 & 10.05 & Yes & \citet{Owers:2012} &\citet{Merlin:2024} \\
         14141 & 00:14:16.6 & -30:23:03 & 0.296 & 9.45 & Yes & \citet{Owers:2012} &\citet{Merlin:2024}\\
         26714 
         & 00:14:08.9 & -30:21:06 & 0.303 & 9.76 & No & Visual check&\citet{Naidu:2024}\\
    \hline
    \end{tabular}
    \tablefoot{Left to right: Galaxy IDs, coordinates (RA and DEC), redshifts, stellar masses, UV coverage (indicating the presence of HST data starting from $\SI{170}{nm}$ rest-frame), RPS identification source, and reference for the coordinates and redshift.}
\end{table*}

\begin{figure*}[htpb]
\centering
\includegraphics[width=0.25\textwidth]{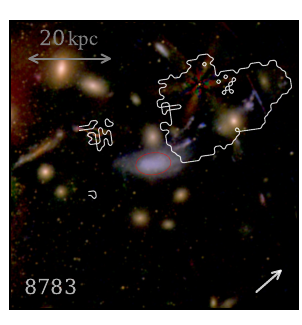}
\hspace{-0.62em}
\includegraphics[width=0.25\textwidth]{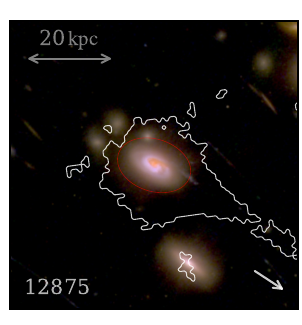}
\hspace{-0.62em}
\includegraphics[width=0.25\textwidth]{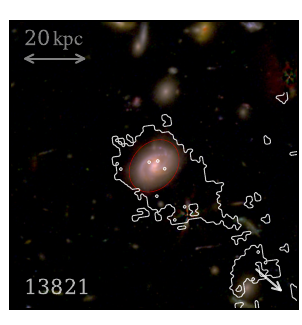}\\
\vspace{-1.2em}
\includegraphics[width=0.25\textwidth]{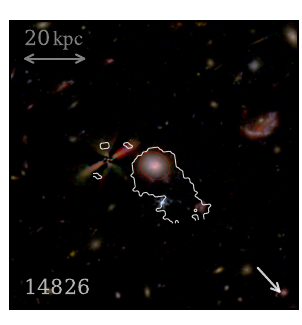}
\hspace{-0.62em}
\includegraphics[width=0.25\textwidth]{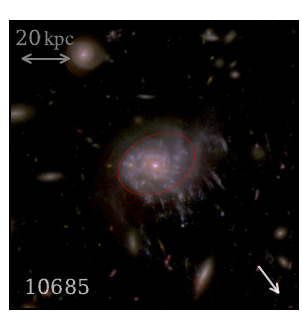}
\hspace{-0.62em}
\includegraphics[width=0.25\textwidth]{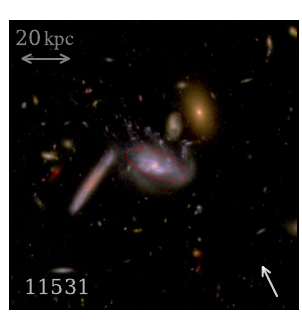}\\
\vspace{-1.2em}
\includegraphics[width=0.25\textwidth]{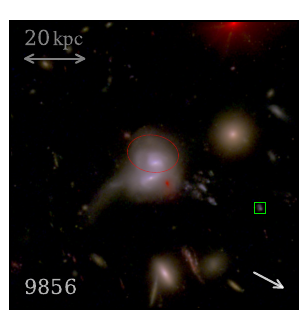}
\hspace{-0.62em}
\includegraphics[width=0.25\textwidth]{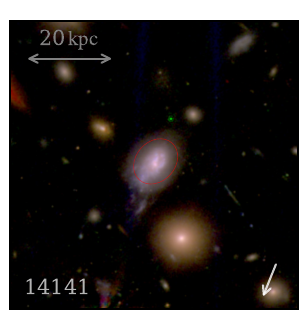}
\hspace{-0.62em}
\includegraphics[width=0.25\textwidth]{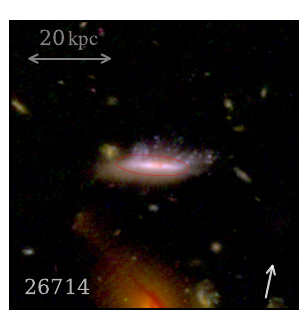}
\caption{Colour composite images of the RPS galaxies analysed in this work, obtained from JWST data. We used F070W, F115W, and F200W filters for the blue, green, and red channels, respectively. The white arrows represent the RPS directions and the red ellipses indicate the galaxy disks, as described in \secref{sec:disk-rps}. The green square in galaxy 9856 indicates a clump with a spectroscopic redshift from MUSE (Astrodeep IDs 9962). In the upper left corner, the physical scale is reported. The white contour is the $\mathrm{H\alpha}$ mask from \citet{Moretti:2020}, when available. The galaxies are organised as follows: first, the RPS galaxies identified by \citet{Moretti:2022}; next, those from \cite{Watson:2025}; and finally, the ones identified by this work. North is up, east is left.}
\label{fig:rgb}
\end{figure*}

\begin{figure}
    \centering
    \includegraphics[width=\linewidth]{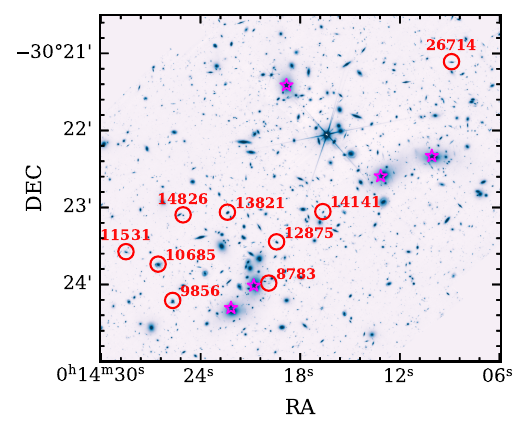}
    \caption{F200W image of the Abell 2744 field. The nine RPS galaxies analysed in this work, labelled with their IDs, are indicated in red. The centres of the dark matter halos from the lensing model of \citet{Furtak:2022} are shown in magenta.}
    \label{fig:map}
\end{figure}

In this work, we focus on the RPS galaxies in the Abell 2744 cluster ($z_{cl}=0.306$), which are shown in Fig. \ref{fig:rgb}. We combine the confirmed RPS galaxies from \cite{Moretti:2022} and \cite{Watson:2025} with two additional candidates from \cite{Owers:2012}, which we confirm as real RPS systems, and include one newly discovered case, for a total of nine galaxies. \tabref{tab:sample} lists the main properties of the sample. To avoid confusion, since most of these galaxies appear in several catalogues with different IDs, throughout this work, we used IDs, coordinates and spectroscopic redshifts from the ASTRODEEP catalogue \citep{Merlin:2024}, when available,  including spectroscopic measurements from various surveys, including UNCOVER \citep{Prince:2024}. Only two galaxies do not have a spectroscopic redshift in this catalogue and we mention them later in this work in the context of their redshift sources.

\cite{Moretti:2022} used VLT/MUSE data to confirm the RPS nature of six galaxies by comparing the ionised gas tail properties to those of their stellar disks. Tails as long as $\SI{10}{kpc}$ were identified. Of these, we exclude two galaxies (A2744-03 and A2774-04 in \citealt{Moretti:2022}) because they are two magnitudes fainter than the other selected galaxies, with F606W magnitudes greater than $25$ \citep[see Table 4 and Fig. 7 of][]{Moretti:2022}, preventing a robust characterisation of their PAH emission. The IDs of the galaxies from \citet{Moretti:2022} in this work are 8783, 12875, 13821, and 14826.

\cite{Watson:2025} studied two additional cluster members (IDs 10685 and 11531), which had previously been marked as potential RPS candidates by \citet{Owers:2012} and \citet{Rawle:2014}, located outside the MUSE pointings. They confirmed RPS instances by investigating the ages of disks and tails through SED fitting based on HST and JWST data \citep[see Sect. 3 of][]{Watson:2025}.

Two more galaxies were identified as RPS by \citet{Owers:2012}: IDs 9856 and 14141. They lack MUSE coverage, but have clear signs of stripping, as also visible from the RGB images and the results of the SED fitting. More details are reported in Appendix~\ref{app:rps-at-play}.

The first object, ID 9856, consists of two merging galaxies at the same redshift, with the galaxies' core spectra identifying them as post-starbursts \citep{Owers:2012}. This system is characterised by two tails. The one extending towards the south-east is quite diffuse, has a curved shape and colour very similar to that of the main galaxies. All these characteristics suggest this tail is most likely the by-product of the merger \citep[see e.g.][]{Mihos:1993, Struck:1999, Vulcani:2021}. In contrast, the western tail is much straighter, more clumpy, and has a much bluer colour. Considering also that one of the clumps in the tail has a spectroscopic redshift compatible with that of the galaxy (see Appendix~\ref{app:rps-at-play} and \figref{fig:muse}), this tail is connected with the two galaxies and is most likely the effect of RPS.

The second galaxy is 14141, a late-type system exhibiting clear asymmetries in both rest-frame UV and optical images. Prominent blue clumps are superimposed on a smoother underlying light distribution, with several extending south-eastwards. Some of these clumps appear detached from the main body, forming a tentative wake.

Here, we present a newly identified RPS galaxy: ID 26714. This galaxy does not have a spectroscopic redshift in the \cite{Merlin:2024} catalogue, but it does appear in the latest release of the ALT survey \citep{Naidu:2024}, from which its redshift was derived. It is an edge-on late-type galaxy exhibiting a striking morphology, with extraplanar material extending northwards. We note that the object appearing north-east of the galactic disk, close to it, has a photometric redshift of $z = 1.23$ and it is therefore likely a background source, unrelated to the cluster environment. Even for this galaxy, more information about its recognition as a RPS galaxy is presented in Appendix~\ref{app:rps-at-play}.

While in most galaxies, the tails seen in JWST imaging are subtle and relatively compact, \cite{Moretti:2022} have shown that all galaxies in their sample exhibit extended H$\alpha$ tails, confirming that RPS is indeed occurring and that the actual extent of the stripped gas may be significantly larger than what is inferred from continuum emission alone. The extent of the H$\alpha$ tails is shown with white contours in \figref{fig:rgb} for the galaxies in the \cite{Moretti:2022} sample.

We note that the galaxy 10685 hosts an active galactic nucleus (AGN), identified through Chandra X-Ray observations \citep{Owers:2012, Watson:2025}. To avoid contamination from the AGN emission, we masked a central circular region with a radius of $\SI{4}{kpc}$, which was excluded from every subsequent analysis. The masked region is broad relative to the point spread function (PSF) of our data, but due to the central source's luminosity, we prefer a conservative approach, as PSF spikes can affect the photometry of the star-forming regions around the galaxy centre and alter SED fitting results. The choice of the mask size does not affect the results of this work, since our study is focussed on local properties, rather than integrated ones. For the remaining galaxies in our sample, there is no evidence of AGN activity \citep{Moretti:2022, Vulcani:2025}. We carefully inspected all the other cluster members in Abell 2744, but we did not detect any other promising RPS galaxies with the data at our disposal. 
The location of the selected galaxies within the cluster is given in \figref{fig:map}.

\section{Methods}\label{sec:methods}

\subsection{Available data}
Abell 2744 is one of the most extensively observed extragalactic fields, with a wealth of space-based data spanning multiple instruments. Observations from HST include WFC3/UVIS, WFC3/IR, and ACS/WFC \citep{Lotz:2017, Treu:2015}, while JWST has added deep imaging and spectroscopy from NIRCam, NIRISS, and NIRSpec \citep{Treu:2022, Paris:2023, Merlin:2024, Mascia:2024, Watson:2025b, Bezanson:2024, Weaver:2024, Suess:2024, Naidu:2024}. Complementing this data, there is also a broad set of ground-based spectroscopic observations \citep{Braglia:2009, Owers:2011, Richard:2021, Prieto-Lyon:2022, Caminha:2019, Bergamini:2023}, providing a wide wavelength coverage. 
Here, we use all publicly available imaging from both HST and JWST. More specifically, we exploited the NIRCam JWST photometry from UNCOVER\footnote{The entire collection of data is publicly available at the \href{https://jwst-UNCOVER.github.io/DR3.html}{UNCOVER website}, which combines all the existing observations \citep{Suess:2024}. The data used in this paper can also be found in MAST: \url{https://doi.org/10.17909/xfym-cg92}} \citep{Bezanson:2024, Weaver:2024} and Megascience surveys \citep{Suess:2024} together with all HST mosaics available on the cluster.
We use the BCGs-subtracted mosaics, when available, which model and then subtract both the intracluster light and many bright cluster members, as described in \cite{Weaver:2024}. The filters F225W, F275W, and F336W do not have BCGs-subtracted images; however, the contribution of the BCGs and intracluster light in these bands is negligible; therefore, we used the non-subtracted images for these filters. The filters F475W and F775W do not cover the galaxies analysed in this study, so they were not used. All together, these surveys cover the central region of the cluster with almost 30 photometric filters from \SI{200}{nm} to almost $\SI{5}{\micro m}$. The large number of filters and the wide range in wavelength allow us to trace the spectral energy distribution (SED) of the galaxies in the cluster with extraordinary detail. The complete list of filters used in this work is given in Appendix~\ref{app:filters}.

\subsection{Alignment and PSF matching}
The wealth of imaging described above is characterised by a different resolution. The first step to use the mosaics was hence to scale down the images to the common resolution of $\SI{0.04}{\arcsec/px}$, which corresponds to $\SI{180}{pc}$ per pixel at the cluster redshift. Subsequently, we matched the PSF of all HST and JWST mosaics to that of the F480M filter, using the \href{https://pypher.readthedocs.io/en/latest/}{\textsc{PyPHER}} tool with a regularisation parameter of $10^{-3}$. The PSFs used to compute the matching kernels are the ones provided by UNCOVER, when available, and the ones from \citet{Watson:2025} otherwise. The resulting PSF full-width half maximum (FWHM) is $\SI{4.4}{px}$ ($\SI{0.18}{arcsec}$), which at the cluster redshift corresponds to a resolution of $\SI{850}{pc}$.
 
While the F444W filter is commonly adopted for PSF matching, we opted for F480M instead. This is because F444W was not used at any stage in our analysis  (not in the extraction of PAH fluxes, nor in the SED fitting) due to its contamination by PAH emission, as further discussed in \secref{sec:sed}. We then aligned the images, with subpixel accuracy, using the \href{https://image-registration.readthedocs.io/en/latest/}{\textsc{image-registration}} package \citep[based on][]{Guizar:2008}. The alignment was computed galaxy by galaxy after cropping a $\SI{20}{\arcsec} \times \SI{20}{\arcsec}$ region around the galaxy centre.

\subsection{PAH flux extraction}\label{sec:pah-extraction}
At the cluster redshift, the $\SI{3.3}{\micro\meter}$  PAH emission feature ($\mathrm{PAH}_{3.3}$) falls within the F430M filter. However, because of the underlying continuum, the total flux measured in this filter cannot be fully attributed to the \ptre feature alone. 
To isolate the continuum contribution, we performed a linear interpolation of the spectral flux density $f_\nu$ between the F335M (Pivot wavelength $\lambda_\mathrm{F335M}=\SI{3.362}{\micro\meter}$) and F480M ($\lambda_\mathrm{F480M}=\SI{4.817}{\micro\meter}$) filters, which we assumed to be dominated by the continuum emission only. Then, we subtracted such continuum from the measured F430M ($\lambda_\mathrm{F430M}=\SI{4.281}{\micro\meter}$) flux to obtain the PAH emission. More specifically, we adopted the same approach used in, for example, \citet{Gullieuszik:2023, Lorenz:2025},  expressed as
\begin{align*}
f_{\mathrm{F335M}}&=f_{\mathrm{F335M}}^\mathrm{\,cont},\\
f_{\mathrm{F480M}}&=f_{\mathrm{F480M}}^\mathrm{\,cont},\\
f_{\mathrm{F430M}}&=f_{\mathrm{F430M}}^\mathrm{\,cont}+\frac{F_{\mathrm{PAH}}}{w_\mathrm{F430M}},\end{align*}
where $F_{\mathrm{PAH}}$ is the PAH flux, $f_\mathrm{filter}$ is the spectral density (in Jansky) in the given filter, and $w_\mathrm{F430M}$ is the rectangular width of the F430M filter ($w_\mathrm{F430M}=\SI{228}{nm}$). The latter was computed using the rectwidth method from the \href{https://synphot.readthedocs.io/en/latest/}{\textsc{synphot}} package.

The resulting equation is
\begin{align}
    F_{\mathrm{PAH}} &= w_\mathrm{F430M}\,\bigl(f_\mathrm{F430M} - (1-k) f_{\mathrm{F335M}} - k f_{\mathrm{F480M}}\bigr),
\end{align}
where
\begin{align*}
    k=\frac{\lambda_{F430M}-\lambda_{F335M}}{\lambda_{F480M}-\lambda_{F335M}}=0.632
.\end{align*}
quantifies the position of the F430M filter's pivot wavelength relative to the baseline set by the continuum-dominated F335M and F480M filters, and serves as a weight in interpolating the continuum at the F430M wavelength.

\begin{figure}
    \centering
    \includegraphics[width=1\linewidth]{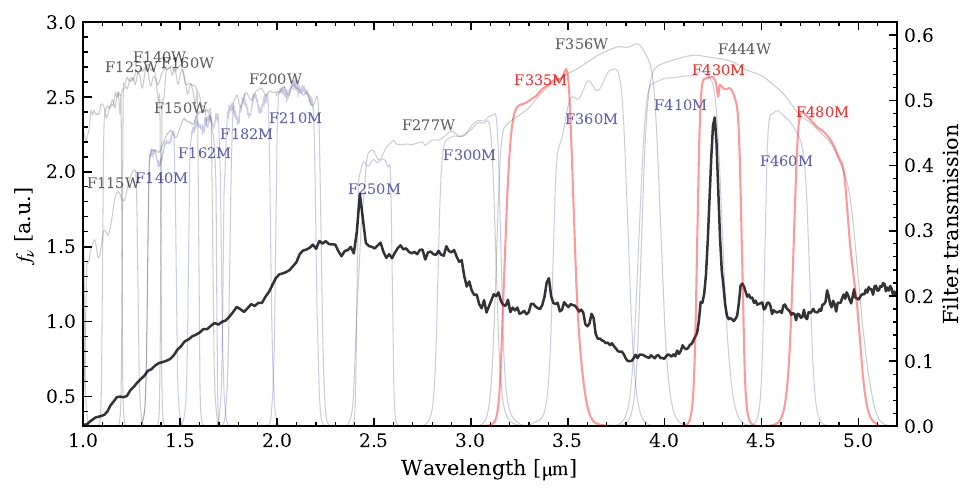}
    \caption{JWST/Prism spectrum of the galaxy 12875 \citep{Vulcani:2025} (black line) and a subset of the JWST and HST filter bandpasses. The complete set of filters used is found in Appendix \ref{app:filters}. The filters used to extract the PAH flux are red coloured: F335M and F480M are used to model the continuum flux. F430M is the filter in which the PAH band falls at the cluster redshift.}
    \label{fig:spectrum}
\end{figure}

Although some filters lie closer in wavelength to F430M than F480M and F335M (potentially providing a better baseline for the continuum interpolation), we chose not to use them for specific reasons (see \figref{fig:spectrum}). We excluded the F410M filter because it is also partially affected by the \ptre emission itself. The F360M filter was avoided due to possible contamination from the broad $\SI{3}{\micro m}$ water absorption feature \citep{Lai:2020}. Similarly, we did not include the F460M filter because it may be affected by the $\SI{3.4}{\micro m}$ aliphatic feature and the PAH emission plateau \citep{Lai:2020, Vulcani:2025}. In Appendix \ref{app:360m-f335m} we report a comparison of the \ptre emission obtained with this selected set of filters (F430M, F335M, and F480M) with the one using the F360M filter instead of the F335M; as in the particular case of the spectrum in \figref{fig:spectrum}, the latter seem to provide a better continuum modelisation. The results obtained with the two sets of filters are consistent, showing a systematic discrepancy in the \ptre flux of 3\%.

We note that galaxies 12875, 14141, and 14826 have NIRSpec/MSA observations of the central regions, and they all show clear \ptre emission \citep{Vulcani:2023, Vulcani:2025}. We specifically measured the equivalent widths of the \ptre emission for these galaxies using an aperture comparable to the MSA shutter used to extract the spectra in \citet{Vulcani:2025}. Our measurements give values between 0.3 and 0.5~dex lower than those computed from the spectra. 
Taking into account various systematic factors, such as the point spread function convolution used in the photometry, slit losses, and the contribution of water absorption bands to the reduction of the continuum, the results are overall comparable. This reassures us of the robustness of the method we used to extract the \ptre fluxes.

We then measured the background noise level ($\sigma_{\rm bkg}$) in the final PAH maps, using the ImageDepth class of \href{https://photutils.readthedocs.io/en/stable/}{\textsc{photutils}}. The background computation was performed locally for each galaxy.  Specifically, we placed 500 circular apertures in regions free from the target galaxy and other detected background sources. This process was iterated three times, applying a 5-$\sigma$ clipping mask to reject outliers. Each aperture had a diameter equal to the PSF FWHM ($\SI{4.4}{px}=\SI{0.18}{arcsec}$), ensuring that the background was sampled at the resolution scale of the data.

Combining the results from all galaxies, we estimated a background-limited $5\sigma_\mathrm{bkg}$ detection threshold in the \ptre maps of $4.8-6.0~\mathrm{nJy}$, corresponding to an AB magnitude limit\footnote{Depth of the image evaluated on a circle of diameter $\SI{0.18}{arcsec}$} of $29.4-29.8$, depending on the source, and to a PAH luminosity of $6-8 \cdot 10^{-18} \mathrm{erg/(s\,cm^2\,arcsec^2)}$.

\subsection{RPS direction and galactic disk}\label{sec:disk-rps}
For each galaxy, we determined the approximate direction of RPS by visually examining the rest-frame optical JWST images and, when available, MUSE data \citep{Moretti:2022}. The direction was inferred from the direction of the stripped material (i.e. the tail). In cases where the tail is not clearly visible in the JWST data, we instead relied on the H$\alpha$ morphology presented in \cite{Moretti:2022}, which is overlaid as white contours in Fig. \ref{fig:rgb}. We note that these contours are only useful to guide the eye and perform qualitative comparisons. The different image depth and PSF size ($\SI{0.58}{\arcsec}$ at $\SI{800}{nm}$, see \citealp{Mahler:2018}) of the instruments used for the observations prevent us from performing a quantitative comparison.

We then measured the extent of each galaxy’s disk in order to distinguish between material still bound to the galaxy and that which has been stripped. Following an approach similar to \citet{Gullieuszik:2020, Moretti:2022}, we developed a procedure based on the F200W image. This band was chosen because it primarily traces the older stellar population and is minimally affected by light from young stars formed during or after the stripping event.
We began by identifying the centre of each galaxy as the centroid of its brightest central region. We then identified the isophote corresponding to a surface brightness level 10$\sigma$ above the background.
The resulting isophotes can appear irregular, particularly along the tail direction. To effectively distinguish between the galaxy disk and the RPS tail, we manually symmetrised the disk. To do this, we selected the side of the galaxy opposite the tail, which is typically less disturbed, and fitted an ellipse to the isophote in that region. We then extended this elliptical shape symmetrically to the more disrupted side of the disk. The resulting contours define a mask that we used to separate the galaxy’s main body from the ram-pressure-stripped tail.
In what follows, we refer to regions within this mask as the galactic disk and those outside it as the tail. The defined disk regions are shown in red in Fig. \ref{fig:rgb}. For galaxy 26714, we adjusted the disk to ensure that its major axis was aligned with the clearly visible galaxy plane (see \figref{fig:rgb}). This adjustment was necessary because the low surface brightness in the south-eastern portion of the galaxy caused the automatic procedure to detect a false large disk that was misaligned with the galaxy plane. 

\subsection{Voronoi binning}\label{sec:voronoi}
To ensure a sufficient signal-to-noise ratio (S/N) when deriving spatially resolved properties of the galaxies (see \secref{sec:sed}), it is necessary to spatially bin the data. 
We performed a Voronoi binning of the images using the F200W band as reference and the \href{https://pypi.org/project/vorbin/}{\textsc{vorbin}} package \citep{Cappellari:2003}, adopting the weighted Voronoi binning algorithm described in \cite{Diehl:2006}. A S/N of 20 was used as the binning target. We selected the F200W band for this purpose. To mitigate potential issues arising from residual PSF mismatches between bands, we imposed a minimum bin size of 15 pixels --- corresponding to the area of a circle with a diameter equal to the  PSF-convolved FWHM. An example of the result of the tesselation on the galaxy 10685 is visible in \figref{fig:astrodendro}: towards the centre of the galaxy, where the S/N is high, the bins tessellate the image with almost hexagonal bins of size equal to the imposed lower limit of 15~px.

\subsection{Clump identifications}\label{sec:astrodendro}
\begin{figure}
    \centering
    \hspace{-3em}
    \includegraphics[width=0.9\linewidth]{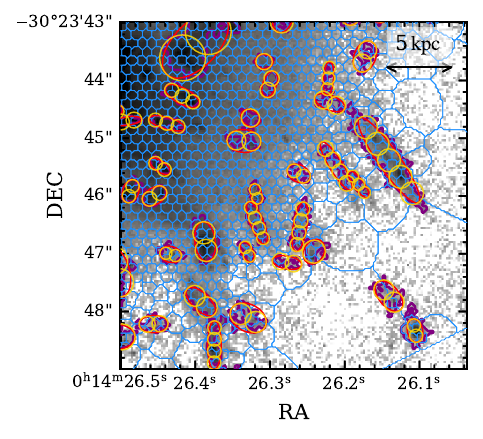}
    \caption{Voronoi binning (in blue) and clumps identification in a portion of the galaxy 10685. The leaves from \textsc{astrodendro} with $N_{min}=30$ are shown in purple; in red, the ellipses traced from the leaves; in yellow, the clumps. Each elongated \textsc{astrodendro} leaf produces a clump association, which is visible as a series of circular clumps. In a clump association, each circle corresponds to a different clump. The scale of the image is shown in the top-right corner. In this image, both the identified clumps inside and outside the galactic disk are visible.}
    \label{fig:astrodendro}
\end{figure}

Star formation occurs in a clumpy fashion, both within the galactic disks and in the RPS tails \citep[e.g.][]{Kenney:2014, Giunchi:2023}. This behaviour is also evident in our sample, as discussed later in this work and which can already be seen in \figref{fig:astrodendro}. For this reason, we next describe the procedure used to identify these clumpy regions, which are then analysed separately from the rest of the galaxies in the following sections.

We identified the clumps using the \href{https://dendrograms.readthedocs.io/en/stable/}{\textsc{astrodendro}} package on the F070W image --- which is the bluest filter among both JWST and HST filters, covering all nine analysed galaxies --- as we expect emission from the star forming clumps to be produced mostly from young stellar populations, with stronger emission towards the bluest part of the spectrum. This Python package constructs a hierarchical structure (dendrogram) from a 2D image, representing regions of contiguous emission as `branches'. New branches form when a saddle point separates two higher-intensity regions, and the smallest substructures in this hierarchy are referred to as `leaves'.

In this work, a single leaf identified by \textsc{astrodendro} does not necessarily correspond to a single physical clump. Observations of RPS galaxies in the Local Universe show that most star-forming clumps in ram-pressure-stripped tails are compact and generally appear as chains of knots, rather than extended filaments \citep{Giunchi:2023}. However, due to the limited resolution in our data (where the PSF-matched angular resolution is $\sim\SI{860}{pc)}$ we were unable to resolve individual clumps. For comparison, \citet{Giunchi:2023} reported typical RPS clump sizes of less than $\SI{140}{pc}$. 

To ensure the identified clumps are significant, we required a new branch to form only if its peak emission exceeded the surrounding region by at least $\delta_\mathrm{min}=3\sigma_{\rm bkg}$ and if it contained more than a minimum number of pixels, $N_{min}$. We adopted $N_{min}$= 30 for all galaxies, except for 26714, which exhibits smaller, more compact clumps; for this galaxy, we used $N_{min}=15$. At the same time, using this threshold for all the galaxies resulted in subdividing regions visible as single structures into smaller, disconnected leaves.

In our analysis, we considered the star-forming clumps in the stripped tails of the galaxies, thereby minimising contamination from disk light. Hence, we removed all the clumps found within the galactic disk, as previously defined. Moreover, for galaxy 9962, in the subsequent analysis, we did not consider the clumps in both the interacting galaxy and the tidal tail, leaving only the ones in the RPS tail (the only case in which we considered these other clumps is in Appendix \ref{app:rps-at-play}).

To explore internal variations within each \textsc{astrodendro} leaf, we divided them into multiple smaller regions. 
For each leaf, \textsc{astrodendro} computed the first and second moments of the pixel distribution, allowing us to approximate the structure as an ellipse (red lines in \figref{fig:astrodendro}). We then converted each ellipse into a series of adjacent circular regions. The radius of each circle is set equal to the semi-minor axis of the ellipse, while the number of circles is determined by the length of the major axis.

In the following, we refer to each individual circle as a `clump'. A set of clumps generated from the same \textsc{astrodendro} leaf is referred to as a `clump association'. All clumps within the same association share the same size, while clumps belonging to different associations may differ in size, depending on the original dimensions of the corresponding leaf.

\subsection{SED Fitting}\label{sec:sed}
We used \href{https://bagpipes.readthedocs.io/en/latest/}{\textsc{bagpipes}} \citep{Carnall:2018} to perform spatially resolved SED fitting for each region of the Voronoi binned image and each clump. We also fitted the integrated photometry in the galactic disk of each galaxy to obtain the stellar masses reported in Table \ref{tab:sample}. To account for the systematic uncertainties in the fluxes \citep[see Appendix C of][]{Watson:2025}, we added a $5\%$ uncertainty in quadrature to the photometry measurements for each filter before conducting the SED fitting. 

\begin{table}[htbp]
    \caption{\centering Parameters used for SED fitting with \textsc{bagpipes}}
    \label{tab:SED-fitting}
    \footnotesize
    \centering
    {\renewcommand\arraystretch{1.3}\begin{tabular}{lcc}
            \hline\hline
            Parameter                                 & Quantity & Prior\\
            \hline
            Redshift, $z$                              & fixed    & Spectroscopic redshift\\                 
            \hline
            Metallicity, $Z$                           & varied   &  $[0.05;1.3]\,\mathrm{Z_{\odot}}$   \\
            Ionization parameter, $\log\mathcal{U}$    & varied   & $[-3;-1]$\\
            Max age of birth clouds, $t_{bc}$            & fixed    & $\SI{10}{Myr}$\\
            \hline
            \multicolumn{3}{c}{\textbf{Dust}}   \\
            Extinction curve model & fixed   &\cite{Calzetti:2000}\\
            Dust emission model & fixed   &\cite{Draine:2007}\\
            V-band extinction $A_V$ & varied &  $[0;3]$ \\
            \hline
            \multicolumn{3}{c}{\textbf{Delayed exponential SFH}}         \\
            Initial mass function                     & fixed    & \cite{Kroupa:2001} \\
            Stellar population model                  & fixed    & \cite{Bruzual:2003}     \\
            Starting SFR age                          & varied    & $[0.002;9.5]\,\mathrm{Gyr}$\\
            Timescale of decrease                     & varied    & $[0.002;9.5]\,\mathrm{Gyr}$\\
            Mass formed                               & varied & $[10^4;10^{10}]\,\mathrm{M_\odot}$\\
            \hline
        \end{tabular}}
    \tablefoot{All the priors are uniform. When two delayed exponentials are used, the parameters described in the SFH section are valid for both.}
\end{table}

The parameters employed in the SED fitting are summarised in \tabref{tab:SED-fitting}. We chose uniform priors for all quantities used in the ranges shown in the table. We fitted each SED with both a single and a double delayed exponential SFH. This choice allows us to capture both simpler evolutionary scenarios, where star formation declines smoothly, and more complex ones, where a secondary burst or a rejuvenation episode might have occurred --- particularly relevant in galaxies affected by RPS, which can trigger or quench star formation in different phases of their evolution. We then calculated the Bayesian information criterion (BIC) estimator \citep{Schwarz:1978} for both results and selected the one with the lowest BIC value. This approach allowed us to use a flexible double delayed exponential when needed, while using only one when two did not significantly improve the fit, based on the BIC value. 
Only 12 out of 205 clumps favour a double delayed exponential SFH and only 3 are isolated, while the other 9 are part of associations where the remaining clumps are fitted with a single exponential. This means no multi-clump association is entirely fitted by the double profile SFH. This finding aligns with the expectation that star formation in the RPS clumps occurs as a single star-forming episode. We therefore forced a single delayed exponential SFH for all clumps to maintain consistency.

Among all the available bands, we excluded F410M, F430M, F460M and F444W from the SED fitting, as these filters are potentially contaminated by PAH or aliphatic emission (see \secref{sec:pah-extraction}). We opted to exclude these bands to ensure a cleaner comparison between physical quantities derived from the SED and those obtained from PAH emission directly. In particular, this approach allows us to compare SFRs inferred from SED fitting and from PAH luminosity in an independent and unbiased manner. We note that the F480M filter was retained in the SED fitting. Therefore, not all filters in the red end of the spectrum were excluded --- only those PAH contaminated --- ensuring we can still robustly constrain the underlying stellar populations.  The complete list of filters used in the SED fitting is given in Appendix~\ref{app:filters}.

The SED fitting results are thus completely independent of the PAH emission. Nevertheless, from the results of our SED fitting, the regions with strong PAH emission have, on average, higher dust extinction. In particular the $A_V$ of regions where the PAH emission have a $\mathrm{S/N} < 3$ is $A_V=0.2_{-0.2}^{+0.6}$ and steadily increase with increasing PAH surface density luminosity $\Sigma_\mathrm{PAH}$, reaching $A_V=0.7_{-0.3}^{+0.4}$ for regions with $\Sigma_\mathrm{PAH}=\SI{1e6}{L_\odot\,kpc^{-2}}$ and $A_V=1.6_{-0.3}^{+0.9}$ for regions with $\Sigma_\mathrm{PAH}=\SI{1e7}{L_\odot\,kpc^{-2}}$. This positive trend confirms that the SED fitting retrieves, on average, more extinction in regions where $\mathrm{PAH_{3.3}}$ is detected, strengthening the SED fitting results about the dust-age degeneracy.

From the SED fitting, run on both the Voronoi bins and the clumps determined in Secs. \ref{sec:voronoi} and \ref{sec:astrodendro}, respectively, we determined the following parameters that are of interest for this work:
\begin{itemize}
    \item The stellar mass surface density, $\Sigma_\mathrm{M\star}$, measured on the sky plane without considering any disk inclination effect;
    \item The surface density of star formation in the on-sky plane, $\Sigma_\mathrm{SFR}^\mathrm{SED}$, averaged over the last $\SI{100}{Myr}$;
    \item The mass-weighted median stellar age ($\mathrm{Age}_{50}$), which represents the age in the history of the studied region at which 50\% of its total mass was formed.
\end{itemize}

\section{Results}\label{sec:results}
We go on to investigate the intensity and spatial distribution of the PAH emission in RPS galaxies to learn how PAH features trace distinct physical processes across the galaxy disks and tails. In particular, we aim to connect the observed PAH emission to key galaxy properties such as stellar mass, star formation activity, and the age of the underlying stellar populations. By doing so, we can gain insight into how the stripping process influences the interstellar medium and recent star formation, and how these effects vary within and among different galaxies in the sample.

\begin{figure*}
\centering
\vspace{-0.7em}
\includegraphics[height=0.25\textwidth]{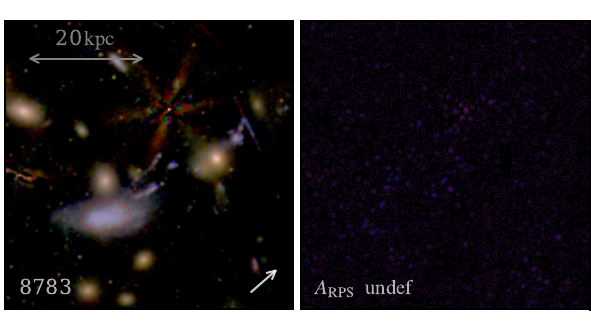}
\vspace{-0.7em}\hspace{0.35em}
\includegraphics[height=0.25\textwidth]{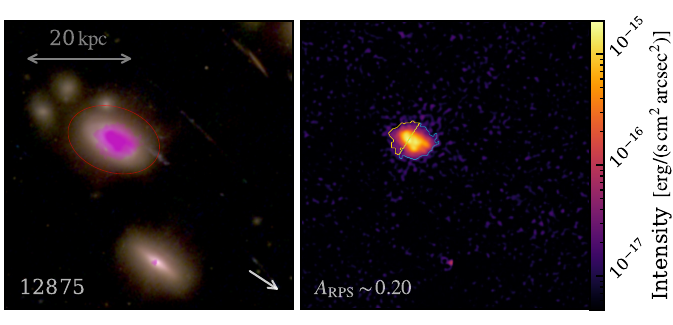}

\includegraphics[height=0.25\textwidth]{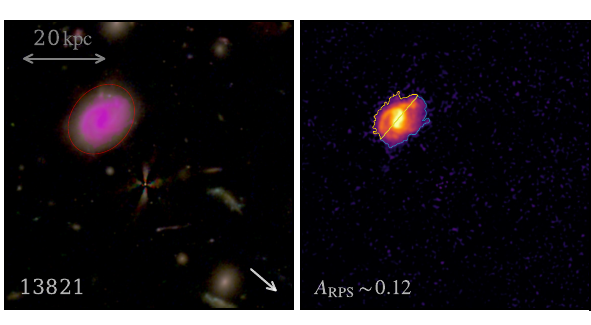}
\vspace{-0.7em}\hspace{0.35em}
\includegraphics[height=0.25\textwidth]{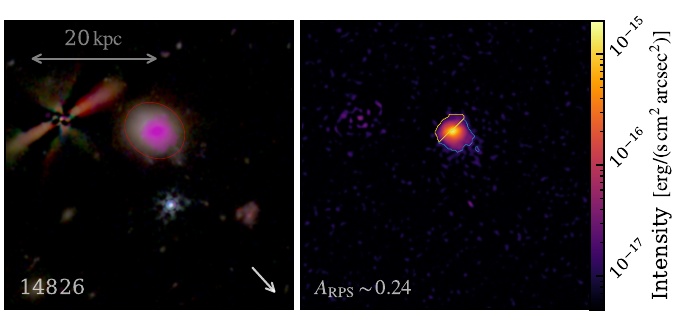}

\includegraphics[height=0.25\textwidth]{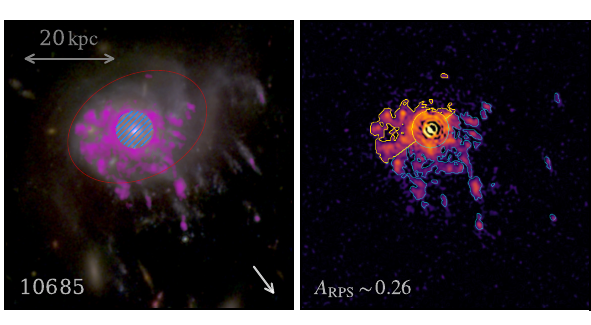}
\vspace{-0.7em}\hspace{0.35em}
\includegraphics[height=0.25\textwidth]{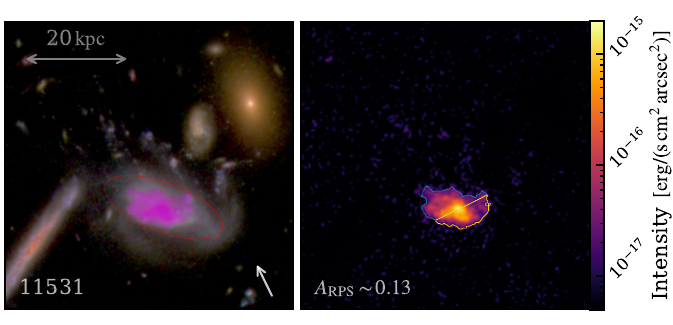}

\includegraphics[height=0.25\textwidth]{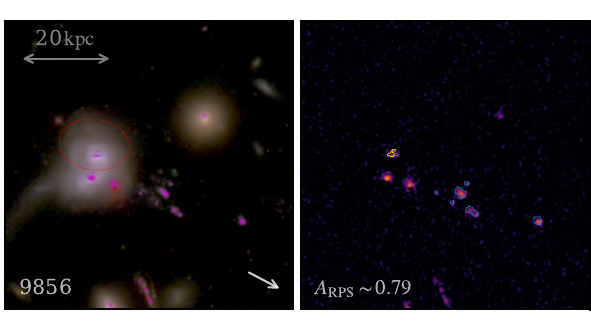}
\vspace{-0.7em}\hspace{0.35em}
\includegraphics[height=0.25\textwidth]{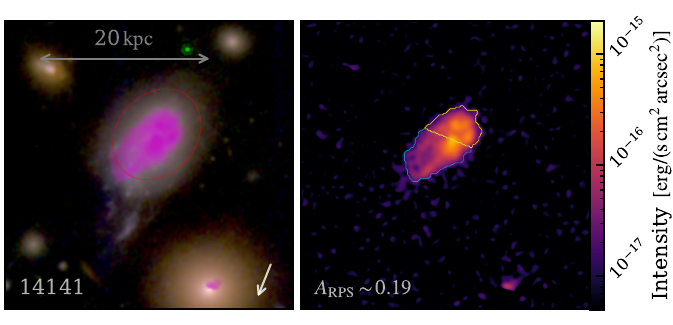}

\begin{center}
\vspace{-0.7em}
\includegraphics[height=0.25\textwidth]{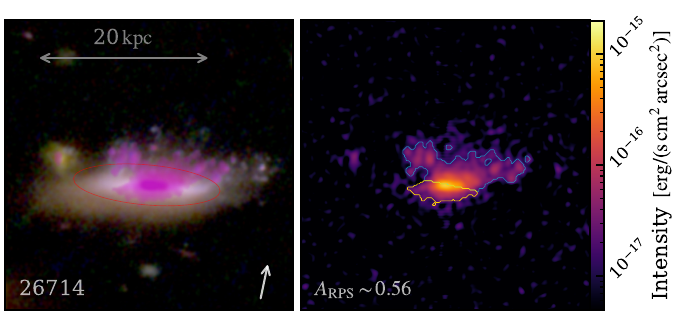}
\end{center}
\vspace{-1em}
\caption{\ptre emission maps for all galaxies in the sample. For each galaxy,  an RGB image composite of F070W (blue), F115W (green), and F200W (red) filters with the $5\sigma$ significance PAH map of the galaxy (superimposed in purple) is shown on the left. The red ellipses indicate the galaxy disks. The arrow indicates the RPS direction. The surface brightness map of the \ptre emission is shown on the right. The gold and blue contours indicate the leading and trailing galactic regions with brightness greater than $5\sigma_{bkg}$, respectively. The $A_\mathrm{RPS}$ is the ram pressure asymmetry of the PAH map as defined in \secref{sec:A-rps}. The blue hatched region in galaxy 10685 represents the masked area. North is up, east is left.}
\label{fig:pah-maps}
\end{figure*}

\subsection{The morphology of PAH emission}\label{sec:pah-morphology}
Figure \ref{fig:pah-maps} illustrates the PAH emission maps of the analysed RPS galaxies, derived as described in \secref{sec:pah-extraction}, overlaid on RGB images for visual reference.  We opted to be conservative, showing only the $5\sigma$ contours of PAH detection in \figref{fig:pah-maps}; however, in this paragraph, we also comment on the $3\sigma$ detection when relevant.

We detected PAH emission in all galaxies except one. The only non-detection occurs in 8783, which shows no PAH signal neither in the disk nor in the tail, even at $3\sigma$ level, implying that the \ptre emission has an upper limit of $\SI{4.3e-18}{erg/(s\,cm^2\,arcsec^2)}$. Interestingly, this galaxy is also the only post-starburst system in our sample \citep{Moretti:2022, Werle:2022, Vulcani:2024}. Despite showing signs of residual star formation in its tail, this activity does not translate into detectable PAH emission. According to \cite{Werle:2022}, star formation in the disk ceased approximately $\sim\SI{10}{Myr}$ ago. The combination of recent quenching and the relatively low stellar mass of the galaxy (\tabref{tab:sample}) likely places its PAH luminosity below our detection threshold. Notably, 8783  also lies closest to the cluster centre in projection, at only $\sim\SI{90}{kpc}$ on the plane of the sky, suggesting it could have experienced the most extreme environmental processing.

Among the remaining galaxies, significant and extended PAH emission is observed. In most cases, the emission is strongly asymmetric; its degree can be quantified, as explained in the following subsection. Here, we focus on the overall visual characterisation of the maps.

Overall, the PAH-emitting regions do not precisely follow the extent of the optical disks and are typically more compact than the stellar boundaries. This observation is not only an effect of the different sensitivity of the \ptre maps, as discussed later in this work. The PAH is predominantly confined within the galaxy disks, although extraplanar emission is evident in some systems. There is no consistent pattern in the morphology, spatial extent, or intensity of the emission.

In galaxy 9856, PAH emission is detected in the very central regions of the two merging components (with a maximum extent of $\lesssim\SI{5}{kpc}$), as well as in the region interpreted as the ram-pressure-stripped tail. Emission is observed as far as $\SI{40}{kpc}$ from the galaxy, in one of the clumps that is spectroscopically confirmed to lie at the galaxy redshift. In contrast, no PAH emission is detected along the tidal tail. A false PAH emitting region is visible at the location of the galaxy ($z_\mathrm{phot}=3.0$) located behind the interacting system. 

Galaxy 13281 exhibits the most regular PAH emission, with a distribution similar to its stellar disk. Galaxies 12875 and 14826 also show fairly symmetric PAH distributions, although the extent of their emission is notably smaller than that of the optical component and the PAH emission centre is off-centred with respect to the galaxy (optical) centre. Galaxy 14826 shows some PAH emission outside the disk in the brightest region of the RPS tail. At the $3\sigma$ level, the entire RGB visible tail is covered by PAH emission, as far as 7~kpc from the galaxy centre.

Galaxy 10685, the most extended system in projection, exhibits a particularly complex and irregular PAH morphology. The emission is fragmented and closely traces the bright blue clumps seen in the RGB image and discussed in \citet{Watson:2025}. As with 9856, PAH emission is detected well beyond the stellar disk along the RPS wake, while no emission is seen along the northern tidal tail --- likely a remnant of a past merger event \citep{Watson:2025}. As noted in Section \ref{sec:sample}, this galaxy hosts a central AGN, and we observe a marked suppression of PAH emission in the central regions. This is consistent with previous findings that AGN activity can destroy PAH molecules \citep[e.g.][]{Smith:2007, O'dowd:2009, Esquej:2014, Sandstrom:2021, Lai:2022}. In contrast, all other galaxies in the sample show central PAH emission, consistent with the absence of AGN activity.

\begin{figure}
    \centering
    \includegraphics[width=0.8\linewidth]{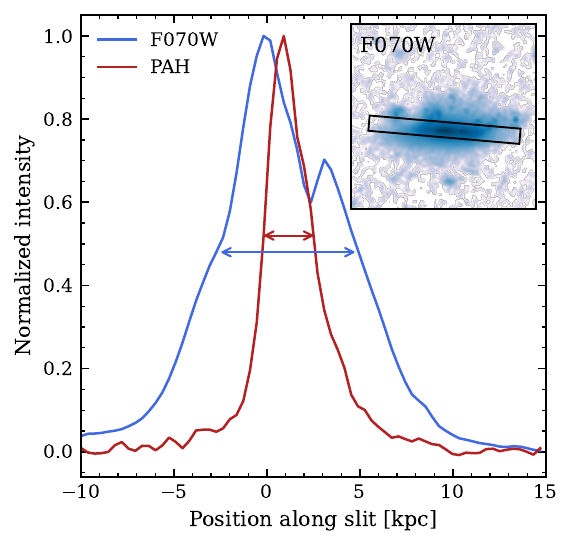}
    \caption{Light profile of the F070W filter and PAH emission for the galaxy 26714, normalised to the maximum luminosity. The horizontal arrows indicate the FWHM of the two profiles. The slit used to compute the light profiles is visible in the top-right panel, overimposed on the F070W map of the galaxy.}
    \label{fig:light-profile}
\end{figure}

Galaxies 26714 and 11531 are notable cases in which the spatial extent of PAH emission is significantly smaller than that of the optical disk. Both galaxies are seen with highly inclined disks (axis ratio smaller than 0.4), with clearly visible tails extending from the disk. Galaxy 26714 shows clear signs of extraplanar PAH emission; similarly, galaxy 11531, while not showing PAH emission in the clumps at $5\sigma$ level, shows emission at $3\sigma$  corresponding to the brightest clumps of the tail, as weakly visible in the right panel of this galaxy in \figref{fig:pah-maps}. In the former, the high inclination angle may facilitate the identification of a distinct truncation in the PAH disk. 
This truncation is quantitatively illustrated by comparing the FWHM of the PAH and F070W ($\sim\SI{540}{\nano\meter}$ rest-frame) light profiles along the galaxy's major axis. Fig. \ref{fig:light-profile} directly shows the light profiles in these two bands for the galaxy 26714.
The light profiles along the slit are calculated using a 3~kpc wide aperture. For 26714, the FWHM of the \ptre light profile is 2.8~kpc, while it is 7.3~kpc in the F200W band. Similarly, for 11531, the values are 2.6~kpc and 5.0~kpc, respectively. Even more interestingly, the FWHM of the light profile of this galaxy in the F435W filter ($\sim\SI{330}{\nano\meter}$ rest-frame) is 5.4~kpc wide. Unfortunately, galaxy 26714 is not covered by any filter in the rest-frame UV part of the spectrum. Such a significant truncation is not observed in other galaxies, which have a FWHM of the PAH that is at least 75\% of the F070W one along the major axis of their galactic disks.

Finally, galaxy 14141 also shows a clear PAH tail. However, the direction of the PAH extension does not perfectly align with the optical tail interpreted as the RPS direction. In particular, blue clumps are visible to the south of the galaxy, whereas the PAH emission extends towards the south-west. The position angle difference between the RPS direction and the PAH elongation is approximately 30°, making this the only galaxy in our sample to show such a misalignment.

The fraction of PAH emitting pixels is approximately 40\% in the galactic disks for all galaxies except for galaxy 8783, which does not have detected PAH emission, and galaxy 9856, which has a fraction of 6\%. The displacement between the centre of mass of the \ptre and the F200W emission map is larger than $\SI{500}{pc}$ only for the galaxies 26714, 11531 and 10685. For the latter one, the result is deeply affected by the masking of \ptre emission in the central region of the galaxy

\subsubsection{Asymmetry of the PAH emission}\label{sec:A-rps}

\begin{table}
    \caption{\centering RPS asymmetry for the galaxies in our sample.}
    \centering
    \begin{tabular}{cccc}
            \hline\hline
          Galaxy ID & $A_\mathrm{RPS}$ & $A_\mathrm{RPS}^\mathrm{min}$ & $A_\mathrm{RPS}^\mathrm{max}$\\
         \hline
         8783& \multicolumn{3}{c}{undefined}\\
         12875 & 0.20 & 0.18 & 0.21\\
         13821 & 0.12 & 0.08 & 0.13\\
         14826& 0.24 & 0.24 & 0.25\\
         10685& 0.26 & 0.17 & 0.36\\
         11531 & 0.13 & 0.09 & 0.18\\
         9856 & 0.79 & 0.78 & 0.82\\
         14141 & 0.19 & 0.17 & 0.21\\
         26714& 0.56 & 0.51 & 0.56\\
         \hline
    \end{tabular}
    \tablefoot{$A_\mathrm{RPS}$ is calculated using the selected RPS direction, using the definition presented in \secref{sec:A-rps}. The minimum and maximum values indicate the range of asymmetry variation when adjusting the RPS direction within a $\pm\SI{15}{\degree}$ range.}
    \label{tab:arps}
\end{table}

As discussed in the previous section, all galaxies have asymmetries in their light distributions.
To quantify the asymmetry in the PAH emission, we defined two regions based on the RPS direction (see \secref{sec:disk-rps}): a leading region, corresponding to the side where the RPS wind impacts the galaxy, and a trailing region, on the opposite side, where stripped material is expected to accumulate. These two regions are separated by a line perpendicular to the stripping direction, passing through the galaxy centre defined as in \secref{sec:disk-rps}. The leading and trailing regions are non-overlapping and together include all pixels with a $\mathrm{S/N}>5$ in the PAH image, even outside the galaxy disk. Contaminants, such as sources with redshifts inconsistent with cluster membership (based on spectroscopic or photometric redshifts), as well as image artefacts, were identified and manually masked. The defined trailing and leading regions are shown in the \figref{fig:pah-maps}, outlined in blue and gold, respectively. 

We define the RPS asymmetry ($A_\mathrm{RPS}$) of the galaxy as
\begin{align}
    A_\mathrm{RPS} = \frac{n_T-n_L}{n_T+n_L},
\end{align}
where $n_T$ and $n_L$ are the number of PAH emitting pixels on the trailing and leading sides, respectively.
$A_\mathrm{RPS}$ ranges between $-1$ and $1$, and $A_\mathrm{RPS}>0$ if there are more emitting pixels on the trailing side than on the leading one.
The $A_\mathrm{RPS}$ measurements are shown on the PAH maps and listed in \tabref{tab:arps}, alongside $A_\mathrm{RPS}^\mathrm{min}$ and $A_\mathrm{RPS}^\mathrm{max}$, which represent the minimum and maximum values obtained when varying the assumed RPS direction by $\pm\SI{15}{\degree}$ to account for the uncertainty in the estimated stripping axis. Overall, varying the angle does not significantly affect the results, ensuring the robustness of the measurement.
All galaxies with detected PAH emission exhibit a positive value of $A_\mathrm{RPS}$, indicating that the \ptre emission is more extended on the trailing side than on the leading side. The asymmetry values range from 0.12 for galaxy 13821 to 0.56 for galaxy 9856, supporting the results from the previous section based on the visual inspection of the PAH maps.

\subsection{PAH emission as a function of local physical properties}\label{sec:scale-relations}
Next, we investigated how the \ptre luminosity correlates with the local physical properties of the galaxies, considering both the Voronoi-binned regions and the identified clumps. Specifically, we considered only the regions emitting in $\mathrm{PAH_{3.3}}$,  comparing the clumps detected in the tails (i.e. outside the galactic disk visible in Figs. \ref{fig:rgb} and \ref{fig:pah-maps}) with the Voronoi bins and discarding all the regions whose overlap with a clump. This selection ensures that the two groups of objects do not have any pixels in common. We note that clumps located within the stellar disks, which are clearly visible in the RGB images, are excluded from both groups. This is because we cannot distinguish between star-forming clumps inside the disk and star-forming ram-pressure-driven clumps that lie on the disk because of the projected view.

\subsubsection{Stellar mass and age}\label{sec:pah-mass}
We examined whether the \ptre luminosity correlates with stellar age and stellar mass. Since clumps and Voronoi bins differ in physical size, we compared the surface density of PAH luminosity and stellar mass, instead of their intrinsic values. This normalisation allows us to compare regions of different sizes consistently, making the analysis independent of the physical extent of the bins or clumps. Moreover, this approach removes the need to deproject the quantities, as both luminosity and mass are affected in the same way by projection effects, under the reasonable assumption that the distribution along the line of sight are comparable for the two quantities.

\begin{figure}[t]
    \centering
    \includegraphics[width=1\linewidth]{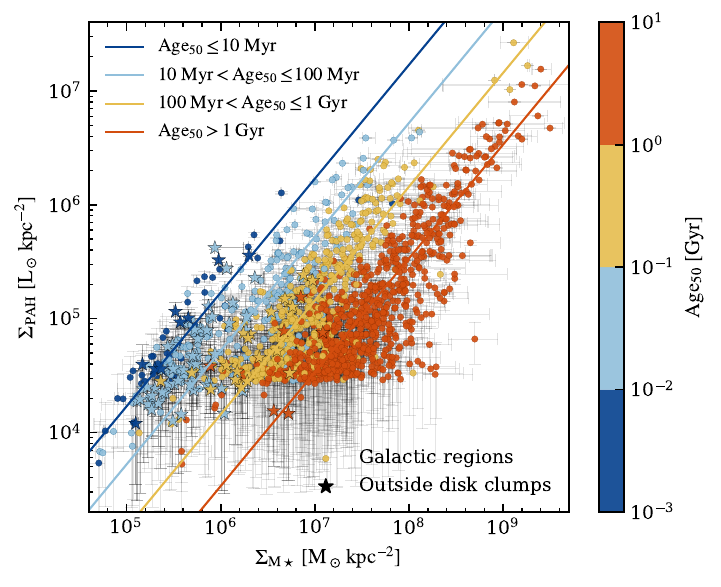}
    \caption{PAH luminosity surface density as a function of stellar mass surface density, colour-coded by the mass-weighted age for each Voronoi cell (circles) and clump (stars). Only the points with an uncertainty smaller than $\SI{1}{dex}$ on each axis are shown in this image. The surface densities are computed on the sky plane, not accounting for deprojection effects. The uncertainties on the $y$-axis are estimated from the background standard deviation (see \ref{sec:pah-extraction}). On the $x$-axis, they come from the SED fitting confidence intervals. The diagonal lines are fit at constant sPAH for each age bin.}
    \label{fig:mass-pah}
\end{figure}
Figure \ref{fig:mass-pah} shows the relationship between the surface \ptre luminosity density ($\Sigma_{\mathrm{PAH}}$) and both the stellar mass surface density ($\Sigma_{M_\ast}$) and the age of the clumps, with the latter two quantities derived from the SED fitting. A correlation between the two quantities is evident, although the scatter is significant and strongly dependent on age. The horizontal cutoff in the PAH luminosity is at around $\sim\SI{2e4}{L_\odot\,kpc^{-2}}$, corresponding to the detection limit for the smallest Voronoi bins ($\SI{15}{px}$ in size), while larger regions and clumps can be detected at lower surface luminosities. 
When controlling for age, the relationship between variables becomes significantly stronger. Specifically, older regions show consistently a lower specific PAH luminosity (sPAH), which is defined as the ratio of PAH luminosity to the stellar mass in each region. In the figure, regions with the same sPAH lie along unity-slope lines. Within each age bin, all data points cluster closely together within a narrow range of sPAH luminosity. In fact, when calculating the fit at constant unitary slope --- which means constant sPAH --- for the data in each log-spaced age bin, ranging from 1 Myr to 10 Gyr, the intrinsic scatter within each bin is found to be between 0.2 and 0.3 dex. sPAH  spans roughly from  $\SI{2e-1}{L_\odot/M_\odot}$ for regions younger than $\SI{10}{Myr}$ down to $\SI{3e-3}{L_\odot/M_\odot}$ for those older than $\SI{1}{Gyr}$. The result of the fit for each age bin is shown as a solid line in \figref{fig:mass-pah}. 
Most regions with an $\mathrm{Age}_{50}$ older than 1 billion years show a very low current star formation; indeed, of these regions, 87\% have a $\Sigma_\mathrm{SFR}^\mathrm{SED}<\SI{0.01}{M_\odot yr^{-1} kpc^{-2}}$ and 97\% a specific SFR of $\mathrm{sSFR}<\SI{1e-9}{yr^{-1}}$. This means that older galaxy regions can still produce detectable levels of PAH emission, supporting the idea that older stars can still generate PAH emissions \citep{Calzetti:2007, Kennicutt:2012}. However, if the amount of stellar mass stays the same, the PAH emission becomes weaker. Notably, the trend between PAH luminosity and stellar mass observed within the galactic disks also holds for the clumps located outside the galactic disks, suggesting that the PAH properties in the stripped tails are similar to those in the main galaxy bodies.

\subsubsection{Comparing different SFR estimates}\label{sec:pah-sfr}
\begin{figure}[t]
    \centering
    \includegraphics[width=\linewidth]{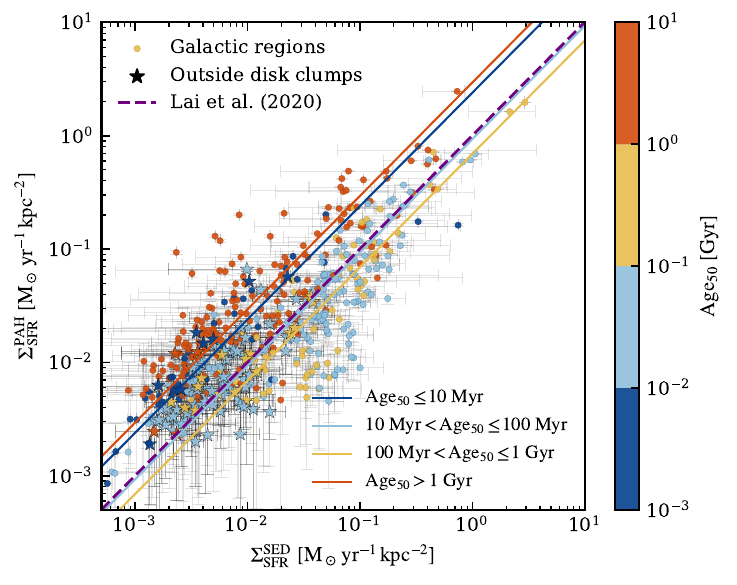}
    \caption{Comparison between the SFR surface density from the SED fitting (on the $x$-axis, averaged on $\SI{100}{Myr}$) and the one inferred from the PAH luminosity ($y$-axis). Points are colour-coded by their mass-weighted median age. The circles represent Voronoi cells that are not overlapped with any clump, while the stars indicate clumps outside the galactic disks. We kept only the measurements for which the uncertainty on the SFR is less than $\SI{1}{dex}$ on both axes. The blue dashed line indicates the 1:1 relation. The coloured lines represent the fit of the data for each age bin.}
    \label{fig:pah-sfr}
\end{figure}
The use of PAH emission features as tracers of obscured star formation has been widely explored in the literature \citep[e.g.][]{Forster:2004, Peeters:2004, Calzetti:2007, Shipley:2016, Xie:2019}. In particular, the \ptre emission has drawn increasing attention, as it is thought to be more sensitive to variations in the local radiation field and more easily destroyed in harsh environments, such as those affected by intense UV radiation or AGN activity \citep{Kim:2012, Lai:2020, Kim:2024}.  Despite this vulnerability, several recent works have shown that \ptre can still be used as a quantitative tracer of star formation, provided that the relevant local conditions are properly taken into account.
However, PAH survival and excitation can be strongly influenced by environmental mechanisms, such as RPS, turbulence, enhanced intra-cluster radiation fields, and shocks; this raises the question of whether PAH features remain reliable SFR tracers in gas that has been displaced or stripped from galaxy disks. The primary goal of this section is therefore to assess the robustness of the PAH–SFR relation in the RPS regime, where the physical conditions differ substantially from those of undisturbed star-forming disks.

We directly compare the SFRs derived from the SED fitting with those inferred from the \ptre luminosity, again in both the Voronoi bins and in the clumps. This comparison allows us to assess the reliability and limitations of PAH-based SFR estimates within different galactic environments, including both the main stellar disks and the stripped tails. We use the relation from \citet{Lai:2020}:
\begin{align*}
\log\biggl(\frac{\mathrm{SFR}}{\mathrm{M_\odot\, yr^{-1}}}\biggr)=-6.8+\log\biggl(\frac{L_{\mathrm{PAH\,3.3}}}{\mathrm{L_\odot}}\biggr).
\end{align*}
We chose to adopt this relation, rather than more recent calibrations focussed on individual star clusters \citep[e.g.][]{Gregg:2024}, since the physical scale of our PSF-matched maps (860~pc) is significantly larger than that of single clusters. Therefore, the physical processes that affect the relation at the single cluster scale are likely not relevant at our resolution.

In \figref{fig:pah-sfr}, we show the comparison between the $\SI{100}{Myr}$-averaged surface SFR density from the SED fitting and the one measured from the PAH luminosity. Overall, the two SFR estimates show good agreement, with most data points lying close to the 1:1 relation, but with a clear age dependence. The overall relation and the age dependence consistency apply to both clumps and regions, including those younger than 10~Myr. Importantly, even though PAH molecules can be partially destroyed by sputtering in the ICM \citep{Micelotta:2010} or excited by shocks \citep{Sivanandam:2014}, our results show that the \ptre emission in RPS clumps still correlates well with the independently derived SFRs. This suggests that, despite the harsher conditions, the surviving PAH population remains sufficiently coupled to the local star-forming activity to serve as a reliable SFR tracer at the physical scales probed here. This reinforces the idea that PAH-based calibrations can be applied (even in RPS environments) provided that environmental effects are considered when interpreting the results. 

A clear trend based on stellar age is evident, as shown by both the data points and the best-fit lines in the figure. The best-fit lines are obtained by fixing a constant 1:1 slope, and the fit takes into account intrinsic scatter. The intrinsic scatter of the relation is measured to be 0.4~dex when fitting all age bins together. However, when fitted separately, the scatter ranges from 0.2~dex to 0.3~dex for all age bins except the oldest one, which has a scatter of 0.4~dex. Regions with intermediate populations ($\SI{10}{Myr}< \mathrm{Age_{50}}\le \SI{1}{Gyr}$) tend to fall near the \citet{Lai:2020} relation. In contrast, both the youngest ($\mathrm{Age_{50}}<\SI{10}{Myr}$) and the oldest regions ($\mathrm{Age_{50}}>\SI{1}{Myr}$) lie systematically above the relation by $0.4-0.5$~dex, suggesting that in these cases the PAH-based SFR might be overestimated. This age dependence aligns with the findings of \citet{Bendo:2020, Zhang:2023}, who identified a similar dependence on specific star formation. This indicates that a higher fraction of evolved stars can shift the relation towards higher PAH values. 

In this section (and the previous one), we report that the \ptre emission in the clumps lie on the same relations as the galactic regions. This means that where we can detect PAH emission, the properties of regions and clumps with the same age are similar. However, to completely understand how the PAH properties of the clumps are comparable to galaxy regions, it is of interest to consider also the PAH non-detections to see if there are clumps where, for example, we can exclude the presence of PAH, while in galactic regions with similar properties we see \ptre emission. The complete analysis is reported in the Appendix \ref{app:pah-age}. We find that the \ptre emission properties of clumps and galactic regions are compatible with each other for every age, even when considering undetected \ptre emission. This compatibility still holds for regions younger than 10~Myr.
Moreover, clumps with an $\mathrm{Age}_{50}$ under $\SI{10}{Myr}$ do not show star formation older than $\SI{30}{Myr}$. Altogether, this indicates that either PAH molecules are stripped from the galaxy or they have formed, spread, and reached equilibrium with the ISM in this timeframe. We offer a broader discussion of these scenarios in \secref{sec:discussion}.

\subsection{Age gradients along the stripped tails }\label{sec:fireball}
\begin{figure}
    \centering
    \includegraphics[width=0.8\linewidth]{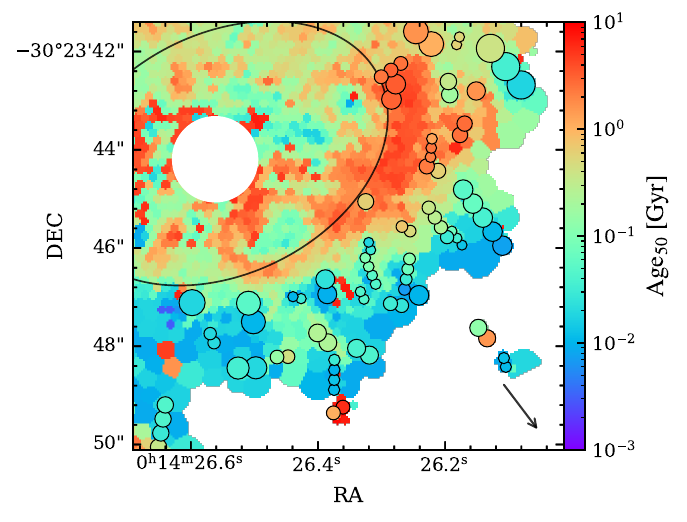}
    \includegraphics[width=0.8\linewidth]{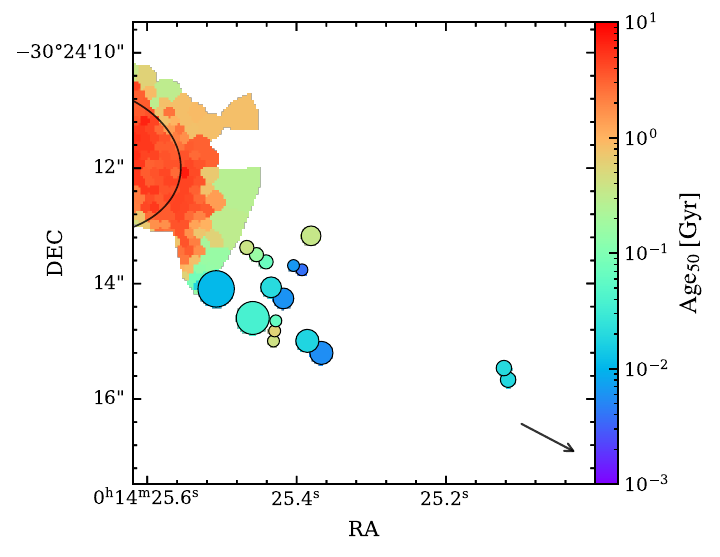}
    \caption{Age estimates from SED fitting for the RPS tails of galaxies ID 10685 (top panel) and 9856 (bottom panel). Circled regions indicate identified clumps. The arrow indicates the stripping direction. The dark contour is the disk ellipse defined in \secref{sec:disk-rps}. The masked central region in galaxy 10685 indicates the $\SI{4}{kpc}$ radius mask centred on the AGN location (see \secref{sec:sed}).}
    \label{fig:age-maps}
\end{figure}
In the previous subsections, we demonstrated that stellar age plays a central role in shaping the strength and detectability of PAH emission across different regions of ram-pressure-stripped galaxies. However, our analysis so far does not directly address the intrinsic age distribution of the clumps themselves, nor whether systematic age gradients are present along the stripped tails. Investigating such gradients is essential to understanding the evolutionary timeline of these clumps: whether they are formed in situ from stripped gas, or whether they originate within the galaxy and are subsequently displaced by ram pressure.
In this subsection, we therefore shift focus from PAH emission to the underlying stellar populations. Specifically, we analysed the stellar ages of the stripped clumps, independently of their PAH luminosity, to uncover spatial trends and reconstruct the formation history of the tails.
\begin{figure}
    \centering
    \includegraphics[width=0.8\linewidth]{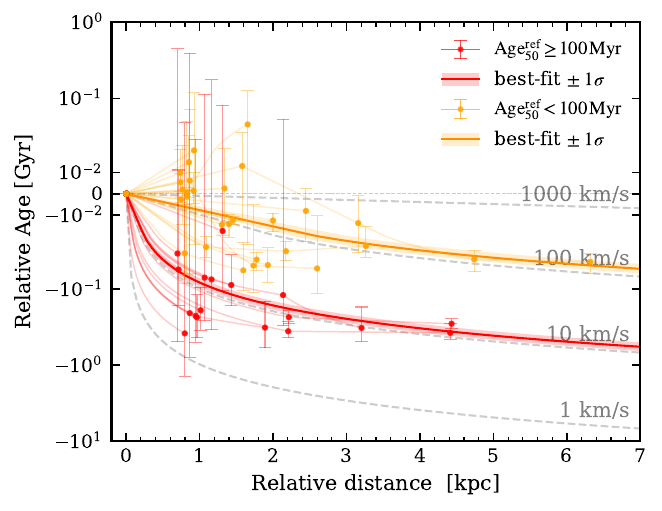}
    \caption{Age distribution of clump associations as a function of relative distance. For each group, the $x$-axis shows the projected distance of each clump from the reference clump (i.e. the one closest to the galaxy within that group). The $y$-axis shows the age of the clumps, from which is subtracted the age of the reference clump of each association. Negative values indicate younger clumps farther away from the galaxy (relative to their group). Associations are divided into two categories: in red, those for which the reference clump is older than $\SI{100}{Myr}$; in orange, those younger than this age. The best fit of each of the two groups with a constant velocity relation is shown as a continuous line. Lines at constant velocity are indicated as grey dashed lines. They do not appear as straight lines due to the symlog scale on the $y$-axis.}
    \label{fig:age-plot}
\end{figure}
\begin{figure}
    \centering
    \includegraphics[width=0.8\linewidth]{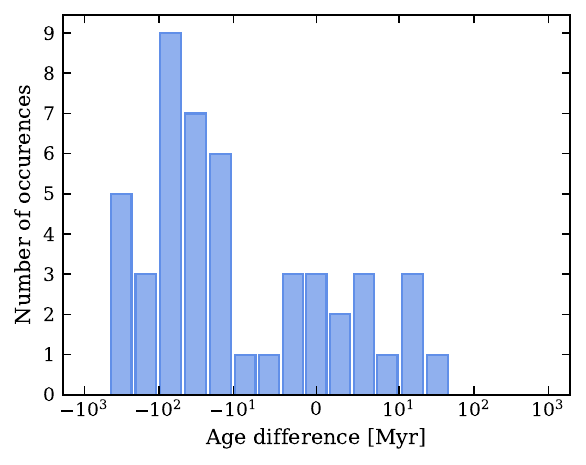}
    \caption{Age differences of consecutive clumps in the same clump association. Negative values correspond to younger ages at greater distance from the galaxy.}
    \label{fig:age-hist}
\end{figure}
Figure \ref{fig:age-maps} presents the stellar age maps of the clumps, along with the Voronoi-binned stellar ages, for galaxies 9856 and 10685, shown as examples. In both cases, many of the clump associations appear well aligned with the RPS direction, and clear age gradients are visible along these structures. The younger clumps tend to lie farther downstream.

For the quantitative analysis described below, we focus on clump associations (as defined in \secref{sec:astrodendro}) that meet the three following criteria: they must must include at least two clumps, lie outside the stellar disks, and have their elongation align with the RPS direction, within a cone of $\pm\SI{30}{\degree}$. This angular constraint accounts for uncertainties in the determination of the RPS direction as well as natural variations in the morphology and orientation of the stripped tails. By selecting clumps with these characteristics, we aim to isolate associations most likely shaped by the effects of the ram pressure, as opposed to internal galactic processes or projection-related artefacts. This approach is deliberately designed to favour purity over completeness: it allows us to maximise the contrast of any physical trends by excluding associations where the signal might be present but weaker or more ambiguous. This choice reflects the fact that the detectability of such age gradients depends strongly on the geometry of the system, including its orientation with respect to our line of sight. While this means we may not capture all possible signatures of RPS-driven evolution across the entire sample, it ensures that the analysed associations are those where the signal is most robust and least affected by noise or contamination.

We analysed how the stellar age varies among the clumps that belong to the same association, independently of the association's distance from the galaxy. For each association, we define the reference clump as the one closest to the galaxy centre in projection. In \figref{fig:age-plot} we show the clump relative age in the association, defined as the age difference between each clump and the reference one, plotted against the projected physical distance\footnote{Since the inclination of the tails with respect to the line of sight is unknown, a full 3D deprojection is not possible.} from the reference clump. \figref{fig:age-hist} further illustrates the distribution of the age difference between consecutive clumps within each association.
Overall, we find that clumps located farther from the galaxy tend to be systematically younger than those closer in, a trend also apparent in the age maps of \figref{fig:age-maps}. This pattern supports the so-called fireball model \citep[][]{Yoshida:2008, Kenney:2014}, in which ram pressure strips and accelerates the gas away from the galaxy as it moves through the intra-cluster medium, while dense gas pockets continue to collapse and form stars downstream. The resulting age gradients trace the relative motion of gas and the newly formed stars during this process. Once stars form, they decouple from the influence of ram pressure and follow ballistic trajectories, whereas the gas that has not yet collapsed continues to be accelerated outwards. Each clump, therefore, provides a snapshot of this ongoing sequence, with its age encoding the timing of collapse along the trajectory of stripped gas.

We modelled the relationship between the relative ages and projected distances of the clumps using a simple linear fit, under the assumption of a constant velocity. This velocity represents the effective speed at which ram pressure separates the uncollapsed gas from the stars newly formed within the clumps. Physically, this can be interpreted as the projected component of the relative motion between the stripped gas and the ballistic stellar component. Several factors contribute to the complexity of this motion, including the galaxy's velocity within the cluster, the relative velocity between the clumps and the galaxy, the density of the ICM at the galaxy's location, and the gravitational potential of the stellar component, which acts to retain the gas. Due to the interplay of these factors (as well as projection effects and possible variations in clump formation timescales), a significant intrinsic scatter around the fitted trend is to be expected.

To investigate possible differences in the dynamical evolution of the clumps, we categorised the associations into two groups based on the age of the reference clump. In Figure \ref{fig:age-plot}, if the clump is younger than 100 million years, it is represented in orange; whereas clumps older than 100 million years are shown in red. 
For the younger group, the  best-fit velocity is $v=\SI{126(18)}{km/s}$, with a small intrinsic scatter of $\SI{5}{Myr}$. In contrast, for the older associations, the inferred velocity is significantly lower, $v=\SI{12.0(1.6)}{km/s}$, with an intrinsic scatter of $\SI{50}{Myr}$.

This result indicates that younger stellar clump associations are more stripped and elongated than older ones (i.e. even a few million years of age difference results in a large physical separation of the clumps). The large scatter for the older group is likely a consequence of both longer dynamical timescales and greater diversity in the properties of their host galaxies, including variations in orbital velocity, ICM density, and viewing geometry. 

Most importantly, the best-fit slopes differ from zero at greater than $5\sigma$ significance, providing robust statistical support for the fireball model scenario. This implies that the sequential formation of clumps, where gas is progressively stripped and collapses to form new stars, plays a key role in shaping the observed age gradients along RPS tails.
\begin{figure}
\centering
    \includegraphics[width=1\linewidth]{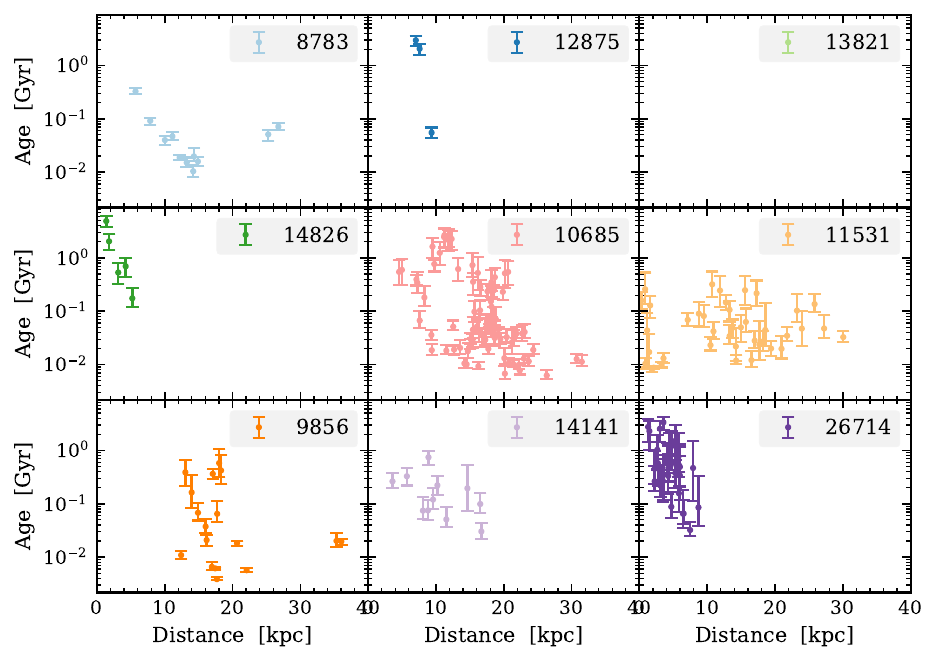}
    \caption{Clump ages as a function of the distance from the galaxy projected along the RPS direction, for each galaxy in the sample.}
    \label{fig:global-age-gradient}
\end{figure}
In addition to the statistical detection of the fireball-model-predicted behaviour within individual clump associations, we also observe a more global age gradient across several galaxies. As shown in \figref{fig:global-age-gradient}, clumps located farther from the galaxy centre tend to be systematically younger than those closer in. This trend is evident, albeit with varying levels of significance and distance range, in nearly all galaxies; the only exceptions are  13821, which lacks detectable clumps beyond the stellar disk, and 11531, where no clear global gradient is detected.

The age gradients observed in the RPS clumps directly affect the distribution of extraplanar PAH emission. Indeed, the PAH luminosity within the clumps depends on the age of the clumps themselves. Specifically, the observed sPAH luminosity decreases from \SI{0.4(1)}{L_\odot/M_\odot} in clumps with a median age of 10~Myr, to \SI{0.05(2)}{L_\odot/M_\odot} at 100~Myr, and down to values as low as \SI{0.005(2)}{L_\odot/M_\odot} for clumps older than 1~Gyr. Consequently, in the context of the fireball model, the peak of the PAH emission is generally located in the downstream region of each clump association where the sPAH is high, but not at the extreme edge, where star formation has just begun and the stellar mass is still too low to produce significant PAH emission. This displacement of the PAH emission is clearly evident in the western-most associations of galaxy 10685, in galaxy 11531, and in several clumps of galaxy 9856 in \figref{fig:pah-maps}.

\section{Discussion}\label{sec:discussion}

In this paper, we demonstrate that RPS significantly affects the spatial distribution of PAH molecules responsible for the $\SI{3.3}{\micro\meter}$ emission, in addition to its well-known impact on the gas content of galaxies \citep[e.g.][]{Koopmann:1998, Jachym:2017, Poggianti:2017}. Specifically, eight out of the nine analysed galaxies (all undergoing RPS) exhibit extended PAH emission. This emission only partially follows the light distribution coming from the stellar disk: we found evidence of both truncated and extraplanar PAH emission.
In particular, we detected PAH emission extending well beyond the galaxy disk, specifically associated with RPS clumps. While extraplanar PAH emission has been previously observed at low redshifts \citep{Kenney:2008, Sivanandam:2014}, and distorted PAH morphologies have been reported in galaxies within the Abell 2744 cluster \citep{Cheng:2025}, this represents the first unambiguous detection of extraplanar PAH emission beyond the Local Universe, with PAH molecules traced in stripped tails up to 40 kpc from their parent galaxy. 

Interestingly, the only two galaxies in our sample that exhibit PAH emission at such large distances from their stellar disks (9856 and 10685) are also the two interacting galaxies. This suggests that environmental pre-processing, such as tidal encounters, could enhance a galaxy’s susceptibility to RPS. In such cases, the removal of gas and PAHs from the disk may be facilitated by gravitational interactions, consistent with findings that pre-processed galaxies are more easily stripped in dense environments \citep{Fritz:2017, Serra:2024, Watson:2025,  Finn:2025}.

In the Local Universe, PAH emission is known to decrease rapidly around young stellar clusters after a few million years \citep{Knutas:2025, Rodriguez:2025, Whitmore:2025}. However, these studies typically probe regions on scales of a few tens of parsecs. In contrast, at the redshift of our sample, the PSF corresponds to a physical scale of approximately 860~pc. On these larger scales, the effects of stellar feedback (e.g. winds and supernovae) in clearing PAHs from cluster surroundings are less dominant, potentially allowing PAH emission to persist longer.

Finally, we tested whether PAH emission remains a reliable tracer of star formation in the context of RPS. Using the calibration from \citet{Lai:2020}, we compared the SFR estimated from the \ptre emission to that derived from SED fitting. The results show good agreement within a scatter of 0.3~dex, for both star-forming clumps outside the galactic disks and for star-forming regions within them. Since the \citet{Lai:2020} calibration was developed for field galaxies, our findings extend its validity to dense environments, suggesting that the \ptre emission remains a robust indicator of star formation — even in stripped clumps affected by ram pressure. But, we found that this relation depends on age, indicating that PAH emission may overestimate star formation in massive, older stellar regions.

\subsection{PAH formation pathways in the context of RPS}

The results shown in this paper raise a fundamental question regarding whether these PAHs are stripped directly from the galaxy along with the gas or whether they form in situ within the stripped clumps.
We started by considering the presence of truncated PAH disks, which suggests that ram pressure can directly deplete these molecules from the outer regions inwards. This outside-in removal is reminiscent of what is commonly observed in the distribution of neutral \citep{Koopmann:1998, Koopmann:2004}, molecular \citep{Jachym:2017, Moretti:2020} and ionised \citep{Poggianti:2017, Gullieuszik:2017, Boselli:2016, Luo:2022} hydrogen during a RPS event. Moreover, the possibility that PAHs themselves are directly affected by ram pressure is physically plausible: PAH molecules are known to be dynamically coupled to the atomic gas phase \citep{Draine:2003, Tielens:2008}, implying that they may be stripped along with the neutral ISM. 

In principle, the observed PAH disk truncation and PAH maps asymmetry (\secref{sec:A-rps}) could result indirectly from the redistribution of the gas due to ram pressure. In this scenario, displaced gas leads to centrally concentrated star formation \citep{George:2025}, and the PAH emission would simply reflect the distribution of young, star-forming regions. However, our observations show that this is not the case, at least for galaxy 11531: its \ptre emission is clearly more truncated than its F435W light (\secref{sec:pah-morphology}), indicating that the \ptre emission does not merely follow the location of the star-forming regions. Indeed, the F435W filter has a rest-frame pivot wavelength at the cluster redshift of $\SI{330}{nm}$, in the UV region, and for this reason directly traces star formation in the last $\SI{100}{Myr}$, similarly to the PAH-tracing timescale (see \secref{sec:pah-sfr}, \citealt{Tielens:2008}). The offset between UV and PAH emission suggests a direct impact of RPS on the PAH molecules themselves, rather than a secondary effect mediated by the star formation distribution.

While our findings support the scenario in which \ptre molecules are stripped by ram pressure, we now consider the alternative possibility that the PAHs observed in the tails of the RPS galaxies are formed in situ. In this context, we aim to identify and discuss the physical constraints that such a formation pathway would imply. Three main mechanisms have been proposed for PAH formation: (i) synthesis in the envelopes of AGB stars; (ii) fragmentation or erosion of larger carbonaceous dust grains; and (iii) the hydrogen abstraction–acetylene addition (HACA) mechanism. In the following, we examine the viability of each of these scenarios in the environment of the stripped clumps. 

In general, AGB stars can contribute to PAH production through their mass loss of carbon-rich material in their stellar winds, where complex organic molecules can form \citep{Frenklach:1989, Cherchneff:1992}. However, our findings strongly disfavour this scenario in the context of the stripped clumps. As shown in Sections \ref{sec:pah-mass}, \ref{sec:pah-sfr}, and Appendix \ref{app:pah-age}, even clumps with median ages below $\SI{10}{Myr}$ already contain \ptre emission comparable to that in similar regions within the galactic disk, despite the latter having had billions of years of ongoing star formation and PAH enrichment.
The formation and release of PAHs from AGB stars requires that a stellar population evolve over at least $\sim$100--300 Myr, depending on the initial mass of the stars involved \citep{Vega:2010, Galliano:2007}. Given that our SED fitting results indicate that star formation in these young clumps began no more than $\SI{30}{Myr}$ ago, there has not been sufficient time for a significant number of stars to reach the AGB phase. While recent starbursts may obscure older stellar populations, the youth of the clumps makes a substantial contribution from AGB stars to the observed PAH content highly unlikely.

Another possible origin of PAHs is the fragmentation \citep{Jones:1994, Seok:2014, Ujjwal:2024} or photo-evaporation \citep{Anderson:2017} of larger carbonaceous dust grains, triggered by shocks or exposure to intense UV radiation. However, this mechanism requires a pre-existing population of dust grains in the ISM. If we assume that the observed PAHs outside the galactic disk are not due to direct stripping, then it follows that larger dust grains, which are generally more resistant to stripping, are also unlikely to have been removed. Therefore, the dust present in the clumps must have formed locally.
On timescales of 10 Myr, dust enrichment is expected to be limited, as Type II supernovae are the major contributors to dust production in the ISM during such early evolutionary stages \citep[e.g.][]{Sarangi:2018, Matsuura:2011, Schneider:2024}. Moreover, the photo-evaporation requires extreme UV fields which, in star-forming regions, are reached only in the local surroundings of newborn stars ($\lesssim\SI{100}{AU}$ \citealt{Anderson:2017}) and the fragmentations from shocks reduce the average size of dust grains, but may destroy the molecular-size grains like the PAHs \citep{Micelotta:2010, Bocchio:2014}. All together, these factors limit the quantity of PAH molecules that can be produced through the evaporation or fragmentation processes, making it unlikely that these processes alone can bring the PAH molecules to the same abundance as in the galactic disk.

Finally, PAH molecules can form through the accretion of smaller hydrocarbon species via the so-called HACA mechanism \citep{Kislov:2013, Yang:2017}. In this process, PAHs grow progressively as acetylene (C$_2$H$_2$) molecules attach to carbonaceous radicals, with hydrogen abstraction steps enabling further additions. This mechanism is particularly efficient in warm ($T\geq\SI{1000}{K}$), dense, and carbon-rich environments, such as circumstellar envelopes or shocked regions, where sufficient concentrations of acetylene and reactive radicals are present. In the context of RPS clumps, invoking the HACA mechanism necessitates the presence of these precursor species. Therefore, even in this scenario, some form of prior stripping from the galaxy is required—either of the small hydrocarbons themselves or of gas enriched enough to produce them rapidly. While such precursors could, in principle, form locally within the clump’s ISM via gas-phase reactions triggered by star formation and UV irradiation, this process is both chemically complex and relatively slow. Indeed, chemical models suggest that PAH formation through HACA in shocked interstellar regions or photodissociation regions occurs on timescales of at least several tens to hundreds of Myr (e.g. \citealt{Micelotta:2010, Slavin:2015}), depending on the physical conditions. The few million years of star formation observed in some of the PAH emitting clumps of this work are insufficient for the full formation of PAH molecules from scratch.

In summary, although dust grain fragmentation may partially contribute, the in situ formation timescales and physical prerequisites make these processes unlikely to fully account for the observed PAH content in the young, stripped clumps.  Our findings thus disfavour in situ formation as the dominant origin of the observed PAHs in the RPS tails and strongly favour the interpretation that PAHs are removed directly by ram pressure along with the gas.

The survival of PAHs in the harsh intracluster environment remains a key question, as the stripped PAHs may evaporate in the interaction with the ICM, destroyed by strong shocks or by the diffuse cluster radiation. However, the PAH detection up to tens of kiloparsecs from the galaxy disk suggests they are not destroyed during the stripping process. Thanks to the \ptre coupling with the atomic gas \citep{Draine:2003, Tielens:2008}, the same physical mechanisms that protect stripped gas from heating and dissociation (e.g. magnetic shielding, suppression of thermal conduction, and confinement in dense clumps) might also protect PAHs \citep[e.g.][]{Tonnesen:2009, Ruszkowski:2012, Gullieuszik:2017, Moretti:2020, Gronke:2020, Muller:2021}.

\subsection{Alternative scenario to explain the age properties of the clumps}
In this work, we also explored the age properties of the clumps, finding that younger clumps are located further down in the RPS tail. Moreover, the clumps are grouped in elongated structures (referred to here as clump associations) and each association present an age gradient within it, with younger clumps further away from the galaxy. This result is compatible with the fireball model \citep{Yoshida:2008, Kenney:2014}. We also find that older clump associations overall have less pronounced fireball effects. Indeed, at constant age difference, younger associations are longer.
The observed galaxy-wide gradient reinforces the interpretation that RPS drives the sequential formation of clumps along the stripped tails. As gas is pushed farther out by the ICM, it requires more time to reach larger distances, and therefore collapses and forms stars at later times. 

Another possible interpretation is that the observed global age gradient can be explained if the gas stripped at earlier times from the outer part of the galaxy (being more violently accelerated and more turbulent, as the galaxy is more diffuse and the gravitational bound weaker) requires longer to cool and condense before forming stars \citep[e.g.][]{Tonnesen:2012, Vijayaraghavan:2017}. Conversely, gas stripped later, from deeper disk regions, is less perturbed by the ICM, allowing earlier collapse closer to the galaxy. As a consequence, older clumps are found nearer to the disk, while younger ones appear farther downstream. 

This second scenario also explains the different fireball model effects observed in young versus old associations. As the young clusters are the ones more affected by the interaction with the ICM, the fireball model produces longer tails with minimal age differences, while older clumps have minimal effects from the fireball model, and the displacement of subsequent star-forming bursts is minimal.
The presence of older clumps nearer to the disk and progressively younger ones farther out is thus consistent with a scenario in which star formation in the tails is not instantaneous, but instead unfolds gradually as the stripped gas moves away from the galaxy.

\section{Conclusions}\label{sec:conclusions}
Using the JWST photometric mosaics from the UNCOVER and Megascience surveys \citep{Bezanson:2024, Suess:2024}, together with ancillary HST data, we mapped the \ptre emission and age properties of the nine most striking RPS galaxy members of the Abell 2744 cluster. At the cluster redshift, the \ptre feature falls within the JWST/NIRCam F430M filter, which captures both the PAH emission and the underlying continuum. To isolate the PAH contribution, we modelled the continuum in this band by linearly interpolating between the continuum-dominated F335M and F480M filters, and subtracted it from the F430M flux. This procedure yielded the \ptre emission maps of the selected galaxies. We analysed the distribution of PAH emission across the galaxies. We also detected clumps beyond the galaxy disk, hosting stars formed from the condensation of the stripped gas. To investigate these systems, we performed a spatially resolved analysis, applying SED fitting techniques to both the galactic disk and the RPS clumps, avoiding PAH-bearing filters.
The findings of this work can be summarised as follows:
\\ \vspace{-1em}
\begin{itemize}
    \item Out of the nine galaxies analysed, eight show \ptre emission at a  $5\sigma$ significance level. The only non-detection corresponds to a low-mass post-starburst galaxy. In two galaxies, the \ptre emission is markedly truncated with respect to the stellar disk, while three galaxies exhibit PAH emission in RPS-induced star-forming clumps located beyond the disk. Moreover, all eight detections display asymmetric PAH emission aligned with the direction of the RPS.
    \item The spatially resolved SED fitting-derived SFR (averaged over the last $\SI{100}{Myr}$) is broadly in agreement with the estimate from the PAH luminosity using the calibration from \cite{Lai:2020}, with an intrinsic scatter of $\SI{0.4}{dex}$. However, in regions with stellar ages younger than $\SI{10}{Myr}$ and older than $\SI{1}{Gyr}$, the PAH-based SFR systematically overestimates that inferred from SED fitting of 0.4-0.5~dex.
    \item The relations observed between PAH emission, SFR, and stellar mass in the disk are consistent with those in the stripped clumps. This consistency still holds for clumps that have only formed stars in the last 30~Myr.
    \item The clumps located beyond the galactic disks are frequently arranged in elongated structures aligned with the RPS direction. Their age gradients, with younger clumps located further away from the galaxy disks, are consistent with the so-called fireball model \citep{Yoshida:2008, Kenney:2014}. 
    The gradients are shallower, implying a higher relative velocity between gas and stars, in overall younger associations. We measured average velocities of  $v\sim\SI{130}{km/s}$ for associations younger than $\SI{100}{Myr}$, compared to $v\sim\SI{12}{km/s}$ for the older ones.
    \item We found a global age gradient in seven of the nine galaxies analysed, with star-forming regions located farther from the galaxy disks being systematically younger on average.
\end{itemize}

In conclusion, our spatially resolved analysis of galaxies undergoing RPS at $z \sim 0.3$ demonstrates that PAH abundance is significantly influenced by ram pressure. Notably, the spatially resolved relations between PAH emission, SFR, and stellar mass in the stripped clumps are found to be consistent with those in the main galactic disk. Furthermore, we present the first evidence of the fireball model in RPS tails beyond the Local Universe, finding its impact to be more pronounced in younger stellar clumps, suggesting that the interaction with the ICM is modulated by the time of stripping and the original location of the gas. Future works that would include a characterisation of PAH emission at longer wavelengths and larger samples are necessary to generalise these findings for larger PAH molecules and to better constrain the age-dependent effects of the fireball model across a broader sample of galaxies.

\begin{acknowledgements}
We appreciate the insightful feedback provided by the anonymous referee, which strengthened both the analysis and presentation. We are grateful to Paola Santini for her helpful support in validating the SED fitting results and to Tommaso Treu and Angela Adamo for useful discussions.

This research is based on observations made with the NASA/ESA Hubble Space Telescope obtained from the Space Telescope Science Institute, which is operated by the Association of Universities for Research in Astronomy, Inc., under NASA contract NAS 5–26555. These observations are associated with programs \#11689 (PI: Dupke), \#13386 (PI: Rodney), \#13495 (PI: Lotz / HFF), \#13389 (PI: Siana), \#15117 (PI: Steinhardt / BUFFALO), and \#17231 (PI: Treu). This work is also based on observations made with the NASA/ESA/CSA James Webb Space Telescope. These observations are associated with programs GO-4111 (MegaScience), MAGNIF (GO-2883, PI: Sun), ALT (GO-3516; PI: Naidu \& Matthee), GO-3538 (PI: Iani), GO-2561 (UNCOVER), ERS-1324 (GLASS), and DD-2767. The raw data were obtained from the Mikulski Archive for Space Telescopes at the Space Telescope Science Institute, which is operated by the Association of Universities for Research in Astronomy, Inc., under NASA contract NAS 5-03127 for JWST. The processed data products were retrieved from the DAWN JWST Archive (DJA). DJA is an initiative of the Cosmic Dawn Center (DAWN), which is funded by the Danish National Research Foundation under grant DNRF140.

P. Watson and B. Vulcani acknowledge support from the INAF Large Grant 2022 `Extragalactic Surveys with JWST' (PI Pentericci) and from the European Union – NextGenerationEU RFF M4C2 1.1 PRIN 2022 project 2022ZSL4BL INSIGHT. P. Watson and B. Vulcani also acknowledge support from the INAF Mini Grant `1.05.24.07.01 RSN1: Spatially resolved Near-IR Emission of Intermediate-Redshift Jellyfish Galaxies' (PI Watson). B. Vulcani and A. E. Lassen acknowledge support from the INAF GO grant 2023 `Identifying ram pressure induced unwinding arms in cluster spirals' (PI Vulcani).

\end{acknowledgements}
\bibliography{aa}

\clearpage
\appendix

\section{Confirming RPS at play in three galaxies}\label{app:rps-at-play}
\begin{figure}[!ht]
    \centering
    \includegraphics[width=\linewidth]{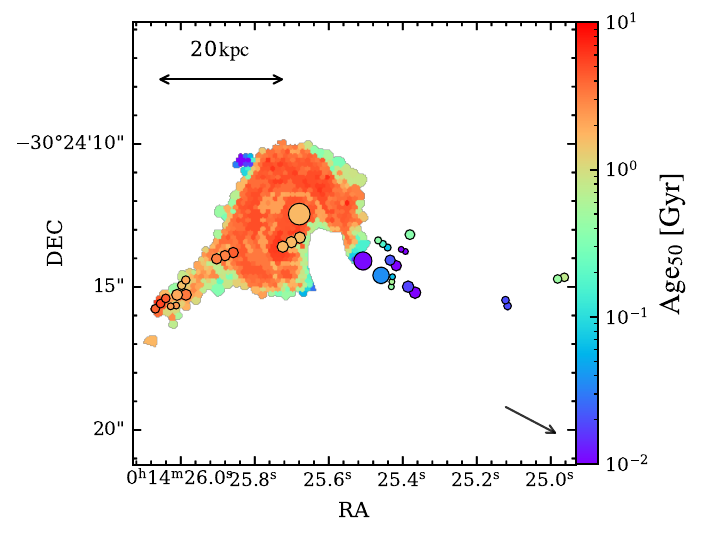}
    \includegraphics[width=\linewidth]{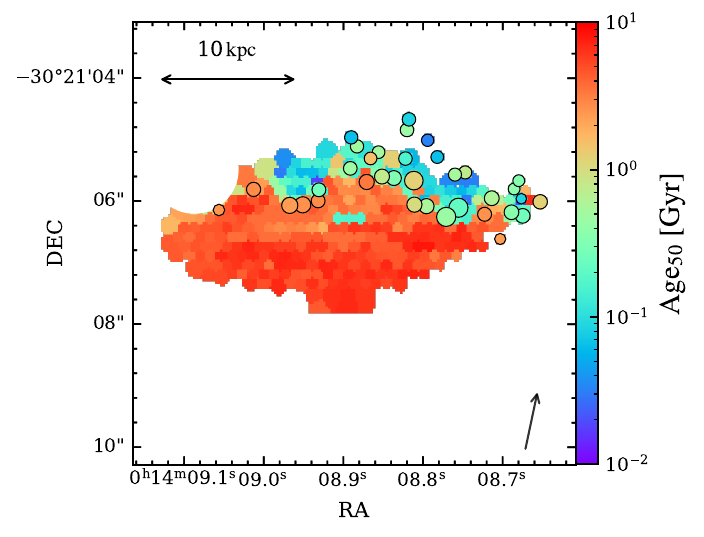}
    \includegraphics[width=\linewidth]{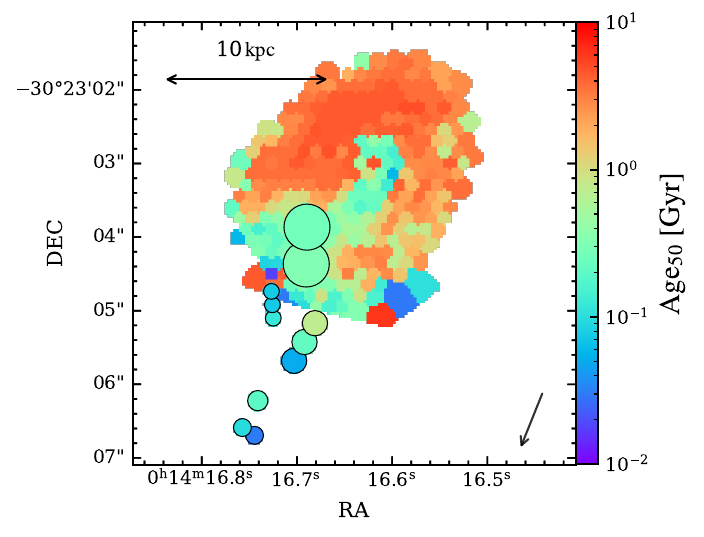}
    \caption{$\mathrm{Age}_{50}$ maps for the galaxies 9856, 14141 and 26714, for both the Voronoi cells and the clumps. This figure includes the clumps located both inside and outside the galactic disk. The arrow indicates the RPS direction. The scale of the image is reported in the upper-left corner. North is up, east is left.}
    \label{fig:ages}
\end{figure}

In this work, we report the identification of one new RPS galaxy (ID 26714) and the confirmation of two tentative RPS galaxies (ID 9856 and 14141, initially classified by \citealt{Owers:2011}) in the Abell 2744 cluster. The classification of the galaxies 14141 and 26714 is based on two main factors: the morphology in the RGB maps (\figref{fig:rgb}) and the mass-weighted age maps obtained from the SED fitting presented in \secref{sec:sed} (\figref{fig:ages}). In particular, in the direction later identified as the RPS tail, both galaxies show clumpy structures extending beyond the galactic disk, associated with recent star formation. The young age of these tails is consistent with what is observed in other RPS galaxies, both in this work and in the literature \citep[e.g.][]{Yoshida:2008, Poggianti:2019}. The combination of morphology, bluer colour and younger ages in the tails compared to the disk supports the interpretation of RPS, which removes gas from the outer regions and triggers star formation preferentially in the central disk and stripped tails. 

Although no spectroscopic data are available for these extraplanar regions, the morphological features strongly suggest the presence of ram-pressure stripping. Overall, the galaxy 14141 bears a striking resemblance to other known RPS systems \citep[e.g.][]{Moretti:2020, Ebeling:2014, McPartland:2016}, supporting its classification as such. The clumpy tail in the galaxy 26714 closely resembles the dust pillars observed in local RPS systems such as NGC 4921 \citep{Kenney:2015}, suggesting a similar physical origin for the observed tail.

The case of the galaxy 9856 is more complex, as it consists of two interacting galaxies. From the RGB map and the age maps, we infer that the south-eastern tail, whose stellar ages are comparable to those of the two galaxy disks, is most likely produced by tidal interaction. In contrast, the south-western tail is clumpier and significantly younger, pointing instead to RPS. 

\begin{figure}
    \centering
    \includegraphics[width=0.8\linewidth]{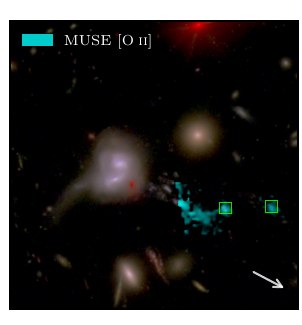}
    \caption{RGB image of the galaxy 9856 composed using the same JWST/NIRCam filters of \figref{fig:rgb} with overimposed, in cyan, the $\textsc{[O\,ii]}$ map from MUSE data (PI: J. Richard). Based on data obtained from the ESO Science Archive Facility with DOI: \url{https://doi.org/10.18727/archive/42}. The tail appears truncated and detached from the galaxy due to the data coverage of the MUSE datacube. In green, the clumps for which the redshift is measured in the Astrodeep catalogue.}
    \label{fig:muse}
\end{figure}

The association of the western-most clumps (Astrodeep ID 9962 and 10065, respectively nearer and further from the main galaxy marked with green squares in Figures \ref{fig:rgb} and \ref{fig:muse}) with the tail of galaxy 9856 is supported by their redshift measurements \citep{Mahler:2018}, which match that of the galaxy, and by MUSE data shown in \figref{fig:muse}. Although the MUSE data (PI: J. Richard) cover only part of the tail, the \textsc{[O\,ii]} map reveals extended gas emission. This emission is particularly strong at the location of clump 9962 but extends further in the RPS direction, linking it to the other identified clumps in the tail. A similar result was reported by \cite{Moretti:2020} for other RPS galaxies in the same cluster.  These characteristics point to a different nature of this tail with respect to the one in the north-east direction, and indicate that it is more likely compatible with an RPS event \citep[e.g.][]{Roediger:2007, Jachym:2017, Boselli:2022, Giunchi:2023, Poggianti:25}

On this basis, we confidently classify this merging galaxy system as being affected also by ram pressure. The coexistence of tidal effects and RPS is also observed for galaxy 10685 \citep{Watson:2025}. 

\newpage

\section{Filters used in this work}\label{app:filters}
We show in Table \ref{tab:filters} the filters used in this work, both for SED fitting and/or PAH extraction. We do not list the filters that do not cover any of the galaxies analysed in this work.

\begin{table}[hb]
    \caption{\centering Photometric observations used in this analysis.}
    \label{tab:filters}
    \centering
    \begin{tabular}{ccccc}
    \hline\hline \\[-0.87em] 
         \textbf{Instrument} & \textbf{Filters} & \textbf{Use} \\[0.12em]
         \hline \\[-0.87em]
         \multirow{2}{*}{HST/ACS} &F225W, F275W, F336W, &  \multirow{2}{*}{SED fitting}\\[0.12em]
          & F435W, F475W, F606W, \\[0.12em]
          &F814W\\[0.12em]\hline\\[-0.87em]
         \multirow{2}{*}{HST/WFC3} & F105W, F125W  & \multirow{2}{*}{SED fitting}\\ 
         &  F140W, F160W \\[0.12em]
         \hline \\[-0.87em]
         \multirow{6}{*}{JWST/NIRCAM}
          & F070W, F090W, F115W, & \multirow{6}{*}{SED fitting}\\ 
         & F140M, F150W, F162M,\\
         & F182M, F200W, F210M,\\
         & F250M, F277W, F300M,\\
          & F335M, F356W, F360M,\\
          & F480M\\[0.12em]
          \hline\\[-0.87em]
        JWST/NIRCAM & F335M, F430M, F480M  & PAH maps\\[0.12em]
        \hline\hline
    \end{tabular}
\end{table}

\section{Filters choice for continuum modelisation}\label{app:360m-f335m}
In this work, we used filters F335M and F480M to model the continuum emission in the F430M filter, as shown in \secref{sec:pah-extraction}. We did not use the filter F360M as it may be affected by a water ice absorption band \citep{Lai:2020}. However, given that also the continuum under the \ptre band may decrease due to this absorption band, F360M may, in some cases, such as in the spectrum shown in \figref{fig:spectrum}, better reproduce the continuum emission in the F430M filter.

For this reason, we re-computed the \ptre emission maps using F360M and F480M filters for the continuum modelling. We followed the same procedure explained in \secref{sec:pah-extraction}. The final equation is
\begin{align*}
    F_{\mathrm{PAH}} = w_\mathrm{F430M}\,\bigl(f_\mathrm{F430M} - 0.449 f_{\mathrm{F360M}} - 0.551 f_{\mathrm{F480M}}\bigr).
\end{align*}
\begin{figure}
    \centering
    \includegraphics[height=0.47\linewidth]{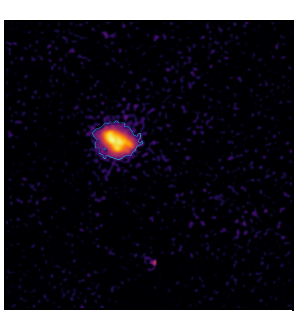}
    \includegraphics[height=0.47\linewidth]{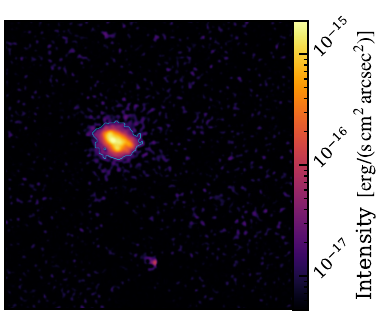}
    \caption{\ptre maps for the galaxy 12875. Left:\ Map using F335M and F480M filters for the continuum modelisation in the F430M filter,  with the same map as in \figref{fig:pah-maps}. Right: \ptre map obtained using F360M and F480M filters to model the continuum.}
    \label{fig:map-comparison}
\end{figure}
\begin{figure}
    \centering
    \includegraphics[width=1\linewidth]{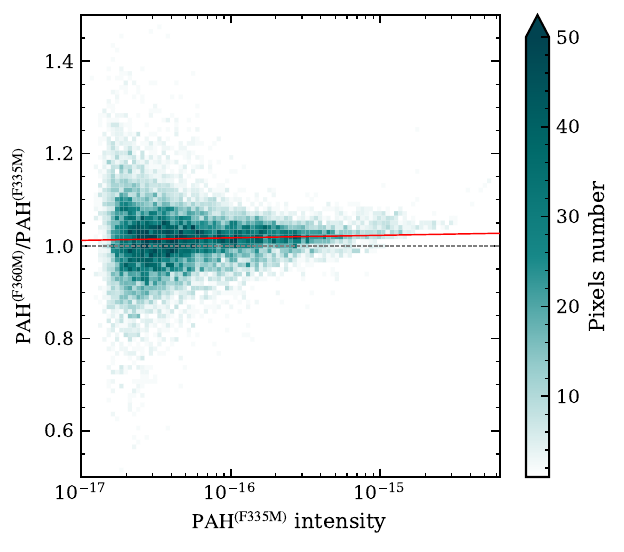}
    \caption{Ratio between the \ptre flux emasured using F360M and F480M vs F335M and F480M for continuum modelisation. The ratio is measured pixel by pixel for all the $5\sigma$ pixels in the maps of the galaxies analysed in this work. On the $x$-axis, the \ptre intensity measured using the F335M-F480M continuum is reported. The dashed grey line is the constant unitary line. The red line is the best fit of the points in the graph.}
    \label{fig:pixel-ratio}
\end{figure}
The emission maps do not have substantial differences in their morphology, as visible in \figref{fig:map-comparison} for one of the galaxies in the sample (ID 12875, the same galaxy for which the spectrum of the central region is visible in Fig. \ref{fig:spectrum}). We manually checked each galaxy, resulting in no major differences: in particular, all the emitting clumps visible in the maps obtained using the F335M filters are still emitting in the map obtained using the F335M filter, and vice versa. 

However, to completely understand how the results differ from each other, we also computed the pixel flux ratio of the \ptre maps obtained using the F360M filters vs the one using F335M, for all the pixels with an emission greater than $5\sigma_{bkg}$ in the F335M case, for all the galaxies. The Figure \ref{fig:pixel-ratio} shows this ratio as a function of the pixel intensity (measured using F335M). The maps obtained using the filter F360M have systematically higher \ptre fluxes --- as visible from the red line in the figure --- of 2-3\%. This systematic behaviour is expected as the F360M flux is generally lower due to the water ice absorption feature, resulting in a lower continuum modelled in the F430M filter and leading to higher \ptre fluxes. The scatter strongly depends on the pixel intensity, starting from $\sim15\%$ at intensities of the order of $\SI{1e-17}{erg/(s\, cm^2\, arcsec^2)}$ to $5\%$ at $\SI{1e-15}{erg/(s\, cm^2\, arcsec^2)}$. 

We note that this scatter is computed on a pixel scale, and the corresponding scatter evaluated on the size of a region or clump is at least a factor of 4 smaller.

\section{Stellar age and the efficiency of PAH emission}\label{app:pah-age}
In Sections \ref{sec:pah-mass} and \ref{sec:pah-sfr}, we considered only those clumps with detected \ptre in emission. We now extend the analysis to fully explore differences between the ram pressure-induced clumps and the galactic disks, including areas where no PAH emission is observed.

To do this, we compare the fraction number of regions classified as strong PAH emitters in the stripped clumps and, separately, of the galactic disks as a function of the $\mathrm{Age}_{50}$. We define a region as strongly emitting if its sPAH luminosity exceeds a certain threshold. For reference, the average sPAH across all regions is $\mathrm{sPAH}>\SI{0.02}{L_\odot/M_\odot}$. We tested different thresholds to prove our results.

Focussing on sPAH luminosity provides a more meaningful comparison than using the PAH luminosity alone. This is because the non-detection in low-mass regions may simply reflect insufficient luminosity to reach the detection threshold, rather than a true physical absence of PAHs. The PAH luminosity is an extensive quantity, as it scales with the stellar mass when any other property, including the SFH profile, is held fixed. Therefore, by normalising for stellar mass, we mitigate biases due to mass differences and better isolate the intrinsic variations in PAH content. Section \ref{sec:pah-mass} and \figref{fig:mass-pah} show that, even with some intrinsic scatter, this scaling property of the PAH luminosity is still valid when considering a fixed $\mathrm{Age_{50}}$ instead of fixing the SFH.

Since the fraction of strong-emitting regions varies as a function of age, we subdivide the regions based on their age into seven logarithmically spaced bins from 1~Myr to 7~Gyr\footnote{We considered the first two bins together, given the low sample size of regions younger than $\SI{5}{Myr}$}. As we cannot directly measure the fraction of strong emitters in each bin, as many regions and clumps have only upper limit detections on the sPAH, we then use the \href{https://lifelines.readthedocs.io/en/latest}{\textsc{lifelines}} package \citep{Lifelines}. This package is designed for survival analysis and is well-suited to handle censored data, which in our context refers to regions where the PAH emission is undetected and only upper limits are available. In our analysis, we consider only those Voronoi bins and clumps with an integrated PAH luminosity detected at a significance level above a $3\sigma$. Regions below this threshold are treated as upper limits in the survival analysis.

To estimate the fraction of strong PAH emitting region in each age bin, we fit the distribution of  sPAH values using the Weibull survival function \citep{Hallinan:1993}, defined as
\begin{align*}
W(x) = \exp\biggl[-{\biggl(\frac{x}{\lambda}\biggr)}^\rho\,\biggr]\quad \lambda > 0,\ \rho > 0,
\end{align*}
where $x$ represents the sPAH and $W(x)$ is the survival function --- that is the probability that a randomly selected region has an sPAH greater than $x$. The parameters $\lambda$ and $\rho$ are determined through maximum likelihood estimation.

The fraction $f$ of strong PAH emitters in each age bin is then calculated as
\begin{align*}
f = 1 - W\left(x = \SI{0.02}{L_\odot/M_\odot}\right),
\end{align*}
namely, 1 minus the value of the survival function evaluated at the threshold.

To account for the uncertainties in the sPAH measurements, we performed a bootstrapping analysis. In each iteration, we perturbed the set of measured sPAH values by incorporating their individual uncertainties, assuming a Gaussian error distribution. The perturbed dataset was then used to fit the Weibull survival model. From the resulting probability distribution function of the survival function, we randomly extracted a value  $f_i$ representing the fraction of strong emitters. We repeated this process multiple times and adopted the final estimate of the fraction, $\hat{f}$, as the mean of the sampled $f_i$ values. The confidence interval on $\hat{f}$ was taken as the range between the \nth{16} and \nth{84} percentiles of the $f_i$ distribution. In addition, we accounted for the uncertainty due to the binomial sampling, which is not included in the confidence intervals provided by the \textsc{lifelines} package. 

\begin{figure*}
  \centering
  \begin{minipage}{0.33\textwidth}
    \centering
    \includegraphics[width=\linewidth]{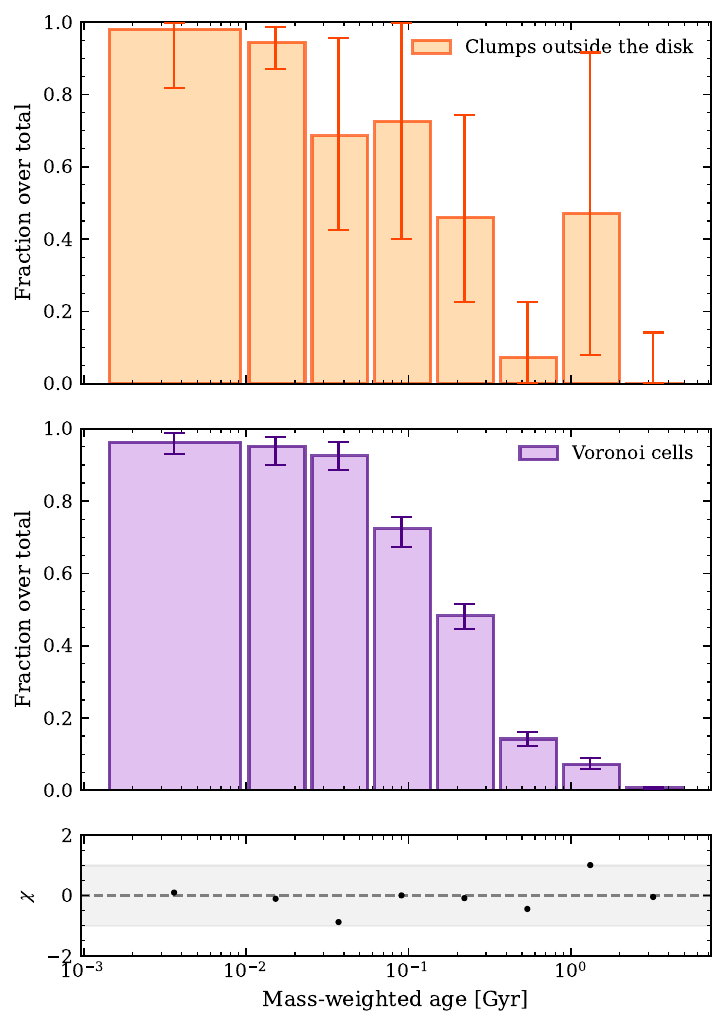}
    \vspace{-0.5em}
    {\small (a) sPAH threshold: $\SI{0.01}{L_\odot/M_\odot}$}
  \end{minipage}\hfill
  \begin{minipage}{0.33\textwidth}
    \centering
    \includegraphics[width=\linewidth]{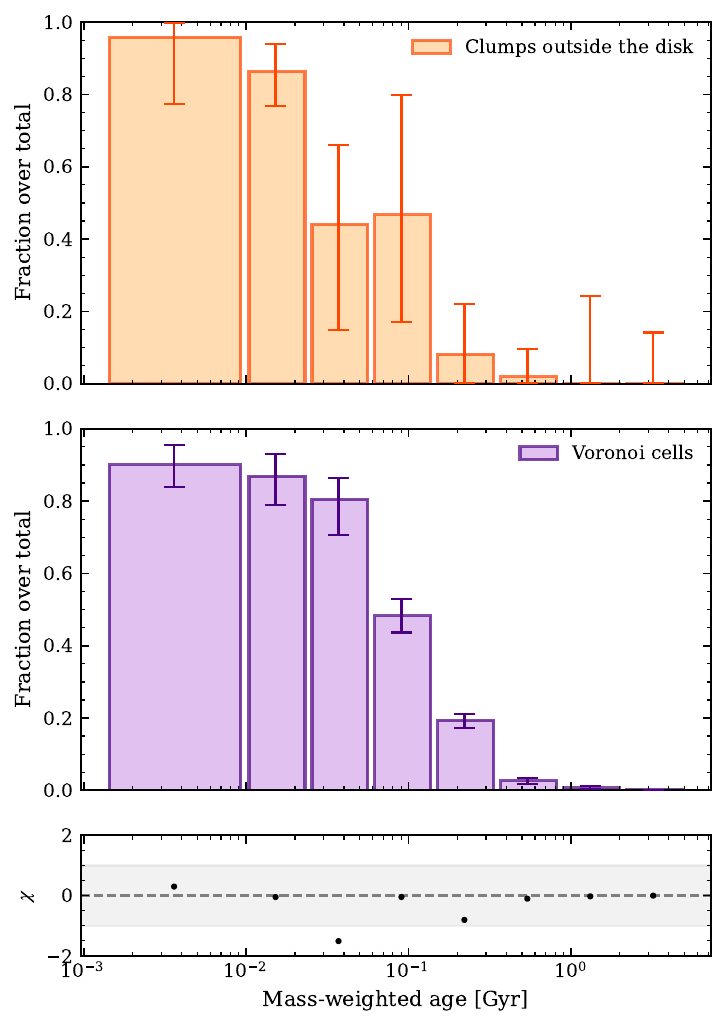}
    \vspace{-0.5em}
    {\small (b) sPAH threshold: $\SI{0.02}{L_\odot/M_\odot}$}
  \end{minipage}\hfill
  \begin{minipage}{0.33\textwidth}
    \centering
    \includegraphics[width=\linewidth]{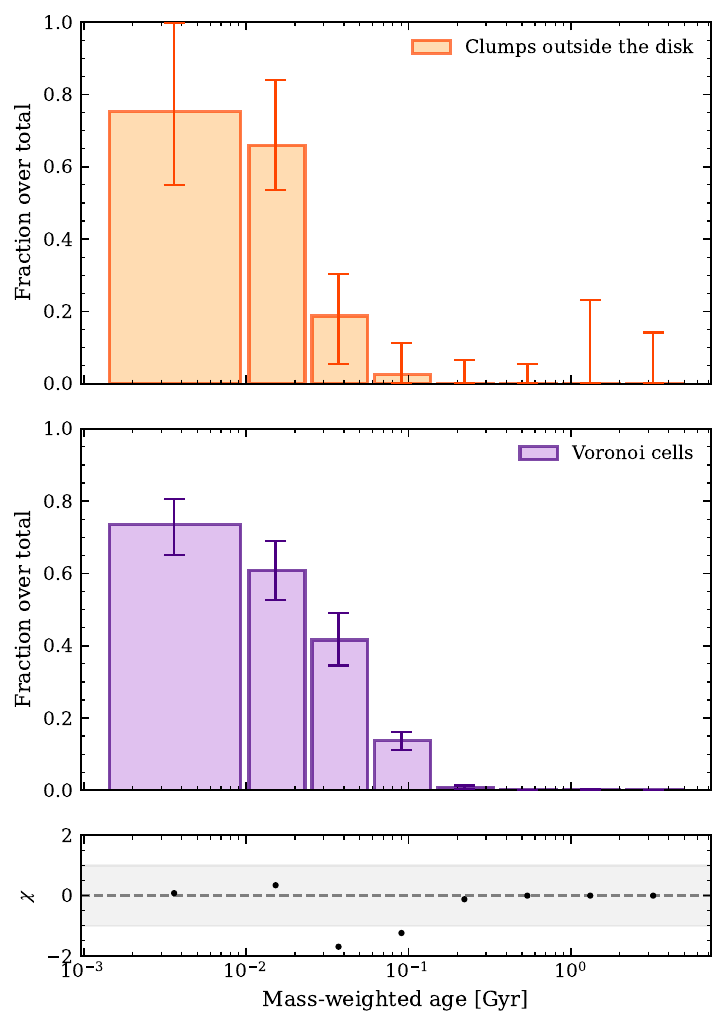}
    \vspace{-0.5em}
    {\small (c) sPAH threshold: $\SI{0.05}{L_\odot/M_\odot}$}
  \end{minipage}
  \vspace{0.5em}
  \caption{Fraction of regions that are strong PAH emitters as a function of the mass-weighted median age. Panels (a)--(c) show the results obtained using different thresholds to define strong PAH emission.  For each threshold, the top panel displays the fraction computed for the clumps in the wake, the middle panel shows the same fraction computed on the Voronoi cells within the galaxy that are not associated with any clump, and the bottom panel highlights the differences between the two distributions, normalised by the uncertainty. 
  The grey shaded area indicates the $1\sigma$ confidence region.}
  \label{fig:pah-age}
  \label{LastPage}
\end{figure*}

Figure \ref{fig:pah-age} shows the fraction of regions classified as strong PAH emitters as a function of the mass-weighted stellar age, or different strong-emitter thresholds. The top panel refers to the stripped clumps, while the bottom one shows the results for Voronoi-binned regions within the galactic disks. 
Although the trend regarding the fraction of strong emitters across different ages is not particularly informative --- since it is influenced by various correlations involving PAH, age, and stellar mass --- a significant observation is that the proportion of strong emitters within the stripped clumps is entirely consistent with that found in the disk regions. This can be seen in the bottom panel of Figure \ref{fig:pah-age}, which illustrates the difference between the histograms based on clump data and those using Voronoi cells. The difference never exceeds $2\sigma$, and all bins, except for one, are compatible within less than $\sigma$. This suggests that the amount and distribution of PAH molecules in the clumps are comparable to what we would expect if they had formed within the galactic disk at the same age.
This agreement is especially significant in the youngest age bin, below \SI{10}{Myr}, with no measurable difference between stripped clumps and star-forming regions within the disk. 

This result further strengthens that the PAH emission in the clumps is compatible with that of galactic regions with the same age and mass, even when we include upper limits. As a result, with the data at our disposal, the non-detection of PAH emission in young clumps is merely a result of the detection limit of our instrument, instead of a lack or depletion of PAH-emitting molecules.
\end{document}